\documentclass[preprint,showpacs,amsfonts,aps,prd,floatfix,superscriptaddress]{revtex4} 
\usepackage{amsmath}
\usepackage{bm}
\usepackage{graphicx}
\usepackage{color}
\usepackage{booktabs}

\usepackage{placeins}
\usepackage{enumerate}

\begin{document}
\title{Real-time quantum dynamics of heavy quark systems\\ at high temperature}
\date{\today}

\author{Yukinao Akamatsu}
\affiliation{Kobayashi-Maskawa Institute for the Origin of Particles and the Universe (KMI), Nagoya University, Nagoya 464-8602, Japan}

\begin{abstract}
On the basis of the closed-time path formalism of non-equilibrium quantum field theory, we derive the real-time quantum dynamics of heavy quark systems.
Even though our primary goal is the description of heavy quarkonia, our method allows a unified description of the propagation of single heavy quarks as well as their bound states.
To make calculations tractable, we deploy leading-order perturbation theory and consider the non-relativistic limit.
Various dynamical equations, such as the master equation for quantum Brownian motion and time-evolution equation for heavy quark and quarkonium forward correlators, are obtained from a single operator, the renormalized effective Hamiltonian.
We are thus able to reproduce previous results of perturbative calculations of the drag force and the complex potential simultaneously.
In addition, we present stochastic time-evolution equations for heavy quark and quarkonium wave function, which are equivalent to the dynamical equations.
\end{abstract}

\pacs{}

\maketitle

\section{Introduction}
Vigorous study of the quark-gluon plasma (QGP) has revealed its strongly coupled nature \cite{QGP}.
This novel state of matter, composed of deconfined light quarks (up, down, and strange quarks) and gluons, which once existed around $10^{-4}$ to $10^{-5}$ second after the Big Bang, is expected to be recreated at the Relativistic Heavy Ion Collider (RHIC) and at the Large Hadron Collider (LHC).
Relativistic hydrodynamics has been used to describe the space-time dynamics of the bulk matter in such collisions \cite{Nonaka:2012qw}, which when compared to elliptic flow data from RHIC revealed a very small shear viscosity ($\eta$) to entropy density ($s$) ratio $\eta/s\approx (1$ - $2.5)/4\pi$ \cite{Song:2010mg}.
The dimensionless ratio $\eta/s$ has since become one of the most sensitive measures of strongly coupled nature of QGP.
In addition to the elliptic flow of light hadrons, which is sensitive to the collective behavior and the initial geometry, other observables such as electromagnetic radiation, jets, and heavy quarks also provide valuable insights to the property of QGP.
Among these probes, we concentrate on the heavy quark probe of QGP.

In relativistic heavy-ion collisions, heavy quarks (charm and bottom quarks) are produced in initial hard partonic processes and are expected to probe the properties of QGP through their subsequent evolution in the medium.
Better theoretical understanding of how the interaction of heavy quarks and medium particles modifies the kinematics and dynamics of the heavy quarks inside the strongly coupled QGP fireball will allow us to utilize them as well controlled hard probes \cite{Muller:2012hr}.
Heavy quarks have been considered to probe transport and static screening properties of the medium.
Momentum spectra of heavy quarks are modified due to energy loss and diffusion processes in the medium \cite{Tywoniuk,Svetitsky:1987gq}.
Survival of heavy quark-antiquark bound states ($J/\Psi$ and $\Upsilon$), or heavy quarkonia, is expected to be sensitive to how the potential is altered by the Debye screening of the color charges \cite{Matsui:1986dk}, although the recombination of heavy quarks into bound states might compensate the possible suppression of heavy quarkonia \cite{BraunMunzinger:2000px}.
Experimental data suggest that all of these phenomena actually take place in heavy-ion collisions at LHC and RHIC.
The suppression of the nuclear modification factor and elliptic flow of open heavy quark observables, namely single electrons, single muons, and $D$ mesons, indicates a strong drag force acting on the heavy quarks \cite{ALICE:2012ab}.
Sequential suppression of $\Upsilon({\rm 1S, 2S, 3S})$ is observed \cite{Chatrchyan:2011pe}, while the recombination processes may be important for $J/\Psi$ \cite{Abelev:2012rv}.
This suggests the usefulness of the $\Upsilon$ family as a clean thermometer of the QGP and also the necessity of knowing wave functions of charm quarks at hadronization.
Motivated by these phenomenological interests, theoretical computations have been performed and are under active development for the diffusion constant \cite{Moore:2004tg,Petreczky:2005nh,Herzog:2006gh}, heavy quark-antiquark free energy \cite{McLerran:1981pb}, spectral function for current correlator \cite{Asakawa:2003re,Laine:2007gj}, and complex potential \cite{Laine:2006ns,Brambilla:2008cx,Rothkopf:2011db,Albacete:2008dz}.
Of these, some \cite{Moore:2004tg,Laine:2007gj,Laine:2006ns} are based on perturbation theory, some \cite{Brambilla:2008cx} on potential non-relativistic QCD (pNRQCD) approach, some \cite{Petreczky:2005nh,McLerran:1981pb,Asakawa:2003re,Rothkopf:2011db} on lattice QCD simulation, and others \cite{Herzog:2006gh,Albacete:2008dz} on the conjecture underlying the AdS/CFT correspondence.
Recent developments in the study of these topics can be found in several review articles \cite{Rapp:2009my}.

In this paper, we will present a new derivation of the dynamical {\it description} of heavy quarkonia in a thermal medium, which might also lead to a better description of the recombination process of heavy quarks.
To discuss the dynamics of heavy quarkonia in the medium and recombination of heavy quarks into bound states, a quantum description is indispensable.
The most convenient framework to study such problems, namely the real-time dynamics of a quantum system interacting with a thermal environment, is that of open quantum system \cite{BrePetText}.
Presently, there exist a few publications based on this approach in similar contexts \cite{Borghini:2011yq,Son:2009vu,Young:2010jq,Akamatsu:2011se}.
In Ref.~\cite{Borghini:2011yq}, analyses using a QCD version of the quantum optical master equation are presented.
It is important to note that the quantum optical master equation is derived under the rotating wave approximation, which assumes the following separation of time scales $\tau_{\rm E},\tau_{\rm S}\ll\tau_{\rm R}$ \cite{BrePetText}.
Here $\tau_{\rm E}$ denotes the correlation time for the environment, $\tau_{\rm S}\approx 1/|\omega-\omega'|$ with $\omega$ and $\omega'$ being the typical frequencies or energy levels of the open quantum system, and $\tau_{\rm R}$ the relaxation time for the open quantum system.
Since the relaxation time of the system scales as $\tau_{\rm R}\propto M/T^2$, this approach should be ideal in studying the real-time dynamics involving transitions among well-separated bound states.
On the other hand, quantum Brownian motion, which is studied in Refs.~\cite{Son:2009vu,Young:2010jq}, is derived by assuming a different time scale separation $\tau_{\rm E}\ll\tau_{\rm S},\tau_{\rm R}$ \cite{BrePetText}.
The spacing of energy levels of the system is expected to get smaller when the heavy quarks are almost liberated from bound states, be it in a heavy-light meson or heavy quarkonium, where thus this approach will become suitable.
Note that quantum Brownian motion of heavy quarks should not necessarily be described by the Caldeira-Leggett model \cite{Caldeira:1982iu}.

We will follow and extend the philosophy of the stochastic potential \cite{Akamatsu:2011se}, which has several appealing features.
The imaginary part of the complex heavy quark-antiquark potential \cite{Laine:2006ns,Brambilla:2008cx,Rothkopf:2011db} can be naturally interpreted as damping due to wave function decoherence between different realizations of the stochastic potential.
It models the random microscopic collisions between heavy quarks and medium particles and thus makes the energy of the heavy quark system fluctuate.
In each realization of the stochastic potential, the particle number of heavy quarks and that of heavy antiquarks are explicitly conserved separately, as they should be in the non-relativistic limit.
In Ref.~\cite{Akamatsu:2011se}, however, it is pointed out that when $M<\infty$ the energy increases linearly without bound in the stochastic description and thus the system does not get thermalized.
What is missing in the stochastic description \cite{Akamatsu:2011se} is an irreversible process, namely friction, which prevents the energy to rise forever.
Remember that not all the forces are described by potentials in the Hamiltonian and a typical counterexample is the drag force.
Therefore when the heavy quark mass is finite $M<\infty$, the proper quantum description of heavy quark systems must incorporate this friction process as well as the screened potential and thermal fluctuations, which are already properly modeled by the stochastic potential in the $M\to\infty$ limit.
Interestingly, even though a classical description is mostly sufficient and consistent in describing energy loss and diffusion processes of single heavy quarks, its quantum description becomes a requisite in describing the quantum dynamics of heavy quarkonia.
A similar argument but in a different context is found in Ref.~\cite{Gallis:1990}

The main purpose of this paper is to study how the in-medium QCD forces, here namely the screened potential, drag force, and their thermal fluctuations, should be incorporated in the quantum description of heavy quark systems from first principle on the basis of the closed-time path (CTP) formalism of non-equilibrium quantum field theory \cite{CalHuText}.
We apply the CTP formalism to QCD and derive the real-time dynamics of heavy quarks as open quantum systems, in leading-order perturbation theory and in the non-relativistic limit.
In this approximation, the effect of the medium enters into the real-time dynamics of heavy quarks through the screening and scattering by the hard medium particles.
As we will see in Sec.~\ref{sec:RTD}, the essence of the real-time dynamics is encapsulated in a renormalized effective Hamiltonian for the doubled degrees of freedom on each time path, which provides all the time-evolution equations of interest: the master equation of the reduced density matrix and the time-evolution equation of the forward correlator for an arbitrary number of heavy quarks and heavy antiquarks.
It is found that thermal fluctuation (friction) derives from the two-point functions of gluon fields in leading order (next-to-leading order) in the low frequency limit and that the order in the low-frequency expansion of the two-point functions directly translates into the order in the non-relativistic expansion of heavy quark interaction terms.
By analyzing these time-evolution equations, the drag force \cite{Moore:2004tg} and the complex potential \cite{Laine:2006ns,Brambilla:2008cx} in leading-order perturbation are reproduced simultaneously.
Concentrating on around the diagonal parts of the reduced density matrix in the position representation, the master equation is reduced to that of the Caldeira-Leggett model \cite{Caldeira:1982iu}.
We will then show how friction can be incorporated in the stochastic time evolution.
We will see that complex noise as well as real noise are necessary in the stochastic description.
The stochastic evolution of the wave function and that of its {\it conjugate}, or in other words the forward and backward propagation of the wave function, turn out to be independent, which is convincing because friction processes are irreversible.

This paper is organized as follows:
In Sec.~\ref{sec:IF}, we will introduce the close-time path formalism of non-equilibrium quantum field theory and derive both the influence functional and the effective action in leading-order perturbation theory and in the non-relativistic limit.
In Sec.~\ref{sec:REH}, we will derive an effective Hamiltonian corresponding to the effective action.
The normal-ordered renormalized effective Hamiltonian is also calculated.
In Sec.~\ref{sec:RTD}, various real-time dynamical equations, namely the master equation for the reduced density matrix, the time-evolution equation of the forward correlator, and a stochastic Schr\"odinger equation, are derived from the renormalized effective Hamiltonian.
In Sec.~\ref{sec:Conclusion}, we conclude our study and give future prospects.
Throughout this paper, we adopt natural units $\hbar=c=k_{\rm B}=1$ and all operators in Hilbert and Fock spaces are denoted by bold fonts.

\section{Influence functional of heavy quarks}
\label{sec:IF}
In this section, we will derive the influence action for heavy quarks with mass $M$ in a medium with temperature $T$ in the leading order of $g$ and up to the order $\sqrt{T/M}$ in the heavy quark mass.

\subsection{Path integral on a closed-time path}
\label{sec:IF-A}
\begin{figure}
\includegraphics[scale=0.3,clip]{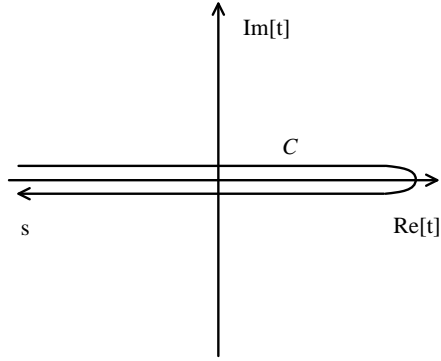}
\caption{
Closed-time path $\mathcal C$ in complex time plane, parameterized by $s$.
}
\vspace{-0.4cm}
\label{fig:CTP}
\end{figure}

In studying real-time dynamics of quantum systems, one often encounters a situation where in the Heisenberg picture $N$-point functions ($N=n+m$) of the form
\begin{eqnarray}
&&{\rm Tr}\left[
{\rm T}\left\{\bm\phi_n(t_n)\cdots\bm\phi_2(t_2)\bm\phi_1(t_1)\right\}
{\bm\rho}
{\rm \tilde T}\left\{\bm\varphi_m(s_m)\cdots\bm\varphi_2(s_2)\bm\varphi_1(s_1)\right\}
\right]\nonumber \\
&& \ ={\rm Tr}\left[
{\rm \tilde T}\left\{\bm\varphi_m(s_m)\cdots\bm\varphi_2(s_2)\bm\varphi_1(s_1)\right\}
{\rm T}\left\{\bm\phi_n(t_n)\cdots\bm\phi_2(t_2)\bm\phi_1(t_1)\right\}
{\bm\rho}\right]
\end{eqnarray}
are of interest.
Here $\rm T(\tilde T)$ denotes the operation of taking the time-ordered (anti time-ordered) product and ${\bm \rho}$ is the density matrix of the system.
In the path integral formalism, such correlators are obtained from the generating functional defined as a functional integration on a closed-time path contour ${\mathcal C}$ labeled by $s$ $(-\infty<s<\infty)$ as illustrated in Fig.~\ref{fig:CTP} \cite{Schwinger:1960qe}.
Correspondingly, source fields in the generating functional must also be defined on the same contour.
In this paper, we assume that the time-evolution of the system starts at $t=-\infty$ and continues forever.
Therefore the contour $\mathcal C$ returns at $t=\infty$ and has both ends at $t=-\infty$.

In QCD at finite temperature, the generating functional is given by
\begin{eqnarray}
Z[\eta,\bar\eta]
&=&\int \mathcal {D}[\bar \psi, \psi, \bar q, q,A]
\langle \bar\psi,\bar q,A \ (s=-\infty)|{\bm\rho}_{\rm tot}|\psi, q, A \ (s=\infty)\rangle\nonumber\\
&& \times\exp
\left[i\int_{\mathcal C}ds \int d^3x\left\{
{\mathcal L}_{\rm QCD}(\bar \psi, \psi,\bar q, q, A)
+\bar \eta \psi+\bar \psi \eta
\right\}\right],
\end{eqnarray}
where $(\eta,\bar \eta)$ are sources for heavy quark fields, $(\bar \psi,\psi)$ are the heavy quark fields, $(\bar q, q)$ are light quark fields, $A=A^{a\mu}$ are gluon fields, and $\mathcal{L}_{\rm QCD}$ is the QCD Lagrangian density.
Here functional integration of the ghost fields and the Faddeev-Popov term are omitted in order to simplify the notation.
For later purposes, we only introduce source fields for heavy quarks.
In the matrix element of $\bm \rho_{\rm tot}$, $|\psi,q,A\rangle$ and $\langle\bar\psi,\bar q,A|$ denote coherent states satisfying for example $\bm\psi(\vec x)|\psi,q,A\rangle = \psi(\vec x)|\psi,q,A\rangle$ and $\langle \bar\psi,\bar q,A|\bar{\bm q}(\vec x) = \langle \bar\psi,\bar q,A| \bar q(\vec x)$.
Since the path integral is obtained by inserting the identity operators made of the coherent states in ${\rm Tr}\left[e^{-i\int_{\mathcal C} ds \bm H_{\rm QCD}}\bm\rho_{\rm tot}\right]$ along the contour $\mathcal C$, the matrix element of $\bm \rho_{\rm tot}$ derives from the insertion at the both ends of the contour ($s=\pm\infty$).
Using the single time representation, this path integral can be rewritten as
\begin{eqnarray}
Z[\eta_1,\bar\eta_1,\eta_2,\bar\eta_2]&=&\int \mathcal {D}[\bar \psi_1,\psi_1,\bar q_1, q_1, A_1]\int \mathcal {D}[\bar \psi_2, \psi_2,\bar q_2, q_2, A_2]\nonumber \\
&& \times\langle\bar\psi_1, \bar q_1, A_1 \ (t=-\infty)|{\bm\rho}_{\rm tot}|\psi_2, q_2, A_2 \ (t=-\infty)\rangle \nonumber\\
&& \times\exp
\Bigl[i\int d^4x\left\{
\mathcal{L}_{\rm QCD}(\bar \psi_1, \psi_1,\bar q_1, q_1, A_1)
+\bar \eta_1 \psi_1+\bar \psi_1 \eta_1
\right\}\Bigr]\nonumber \\
&& \times\exp
\Bigl[-i\int d^4x\left\{
\mathcal{L}_{\rm QCD}(\bar \psi_2, \psi_2,\bar q_2, q_2, A_2)
+\bar \eta_2 \psi_2+\bar \psi_2 \eta_2
\right\}\Bigr]
\end{eqnarray}
with boundary conditions at $t=\infty$:
$\left[\bar\psi_1(\infty)\gamma^0\right]^{\dagger}=\psi_2(\infty)$,
$\left[\bar q_1(\infty)\gamma^0\right]^{\dagger}=q_2(\infty)$, and
$A_1(\infty)=A_2(\infty)$.

In the following, we consider a density matrix in the Heisenberg picture (initial condition of a density matrix in the Schr\"odinger picture) of the form ${\bm\rho}_{\rm tot} = {\bm\rho}^{\rm eq}_{\rm E} \otimes {\bm \rho}_{\rm S}$, where ${\bm \rho}^{\rm eq}_{\rm E}$ is the equilibrium density matrix for the {\it environment degrees of freedom}, i.e. light quarks and gluons, and ${\bm \rho}_{\rm S}$ is an arbitrary density matrix for the {\it system degrees of freedom}, i.e. heavy quarks.
The matrix element of the density matrix is given by
\begin{eqnarray}
&&\langle\bar\psi_1,\bar q_1, A_1 \ (-\infty)|{\bm \rho}_{\rm tot}|\psi_2, q_2,A_2 \ (-\infty) \rangle \nonumber\\
&&=\langle \bar q_1,A_1 \ (-\infty)|{\bm \rho}^{\rm eq}_{\rm E}|q_2,A_2 \ (-\infty)\rangle\cdot
\langle\bar\psi_1(-\infty)|{\bm \rho}_{\rm S}|\psi_2(-\infty)\rangle.
\end{eqnarray}
This density matrix represents an initial condition where the environment and the system are decoupled and the former (latter) is in equilibrium (arbitrary) state.
We are interested in the time evolution of the system degrees of freedom from this initial condition. 

\subsection{Integrating out the environment degrees of freedom}
\label{sec:IF-B}
Since we are interested in the dynamics of heavy quarks as an open quantum system, let us integrate out the gluon and light quark fields:
\begin{eqnarray}
Z[\eta_1,\bar\eta_1,\eta_2,\bar\eta_2]&=&\int \mathcal {D}[\bar \psi_1, \psi_1,\bar \psi_2, \psi_2]
\langle\bar\psi_1(-\infty)|{\bm\rho}_{\rm S}|\psi_2(-\infty)\rangle\\
&&\times \exp\Bigl[i\int d^4x\left\{
\mathcal{L}_\psi(\bar \psi_1,\psi_1)
+\bar \eta_1 \psi_1+\bar \psi_1 \eta_1
\right\}\Bigr]\nonumber\\
&& \times\exp
\Bigl[-i\int d^4x\left\{
\mathcal{L}_\psi(\bar \psi_2, \psi_2)
+\bar \eta_2 \psi_2+\bar \psi_2 \eta_2
\right\}\Bigr]
\cdot Z_{\rm FV}[\bar \psi_1 t^a\gamma^{\mu}\psi_1,\bar \psi_2 t^a\gamma^{\mu}\psi_2],\nonumber\\
Z_{\rm FV}[j^{a\mu}_1,j^{a\mu}_2]&=&
\int \mathcal {D}[\bar q_1, q_1, A_1,\bar q_2, q_2,A_2]
\langle \bar q_1,A_1 \ (-\infty)|{\bm\rho}^{\rm eq}_{\rm E}|q_2,A_2 \ (-\infty)\rangle \\
&& \times\exp
\Bigl[i\int d^4x\left\{
\mathcal{L}_{g+q}(\bar q_1, q_1, A_1)-gj_1^{a\mu}A^a_{1\mu}
\right\}\Bigl]\nonumber \\
&& \times\exp
\Bigl[-i\int d^4x\left\{
\mathcal{L}_{g+q}(\bar q_2, q_2, A_2)-gj_2^{a\mu}A^a_{2\mu}
\right\}\Bigr].\nonumber
\end{eqnarray}
Here $\mathcal{L}_{\psi}$ and $\mathcal{L}_{g+q}$ are the Lagrangian densities of free heavy quarks and interacting gluons and light quarks.
The couplings between heavy quarks and gluons appear in the Feynman-Vernon influence functional $Z_{\rm FV}[j^{a\mu}_1,j^{a\mu}_2]$ \cite{Feynman:1963fq} as source terms for gluons.
Note that $\ln Z_{\rm FV}[j^{a\mu}_1,j^{a\mu}_2]$ is a generating functional for contour-ordered connected $N$-point functions of gluons $(N=n+m)$:
\begin{eqnarray}
&&\left(-\frac{1}{ig}\right)^n\left(\frac{1}{ig}\right)^m
\frac{\delta^n}{\delta j^{a_1\mu_1}_1(x_1)\cdots\delta j^{a_n\mu_n}_1(x_n)}
\frac{\delta^m}{\delta j^{b_1\nu_1}_2(y_1)\cdots\delta j^{b_m\nu_m}_2(y_m)}
\ln Z_{\rm FV}[j^{a\mu}_1,j^{a\mu}_2]\mid_{j^{a\mu}_1=j^{a\mu}_2=0}\nonumber\\
&& \ =\left<
\tilde {\rm T}
\left\{ {\bm A}^{b_1}_{\nu_1}(y_1)\cdots {\bm A}^{b_m}_{\mu_m}(y_m)\right\}
{\rm T}\left\{{\bm A}^{a_1}_{\mu_1}(x_1)\cdots {\bm A}^{a_n}_{\mu_n}(x_n)\right\}
\right>_{\rm conn}.
\end{eqnarray}
Then, if we assume $\langle {\bm A}^{a\mu}(x)\rangle=0$, the expansion of $\ln Z_{\rm FV}[j^{a\mu}_1,j^{a\mu}_2]$ in terms of $j^{a\mu}_1$ and $j^{a\mu}_2$ is
\begin{eqnarray}
\ln Z_{\rm FV}[j^{a\mu}_1,j^{a\mu}_2]
&=&-\frac{g^2}{2}\int d^4x d^4y
\left(
j^{a\mu}_1(x),\ j^{a\mu}_2(x)
\right)\nonumber\\
&&\times\left(
\begin{array}{cc}
G^{\rm F}_{ab,\mu\nu}(x-y)
 & -G^<_{ab,\mu\nu}(x-y) \\
-G^>_{ab,\mu\nu}(x-y)
 & G^{\rm \tilde F}_{ab,\mu\nu}(x-y)
\end{array}
\right)
\left(
\begin{array}{c}
j^{b\nu}_1(y)\\
j^{b\nu}_2(y)
\end{array}
\right)
+\mathcal O(g^3)\\
&\equiv& iS_{\rm FV}\left[j^{a\mu}_1,j^{a\mu}_2\right],\nonumber
\end{eqnarray}
where the (bare) two-point functions are defined by
\begin{eqnarray}
G^{\rm F}_{ab,\mu\nu}(x-y)
\equiv \langle{\rm T}{\bm A}^a_{\mu}(x){\bm A}^b_{\nu}(y)\rangle, && \
G^{\rm \tilde F}_{ab,\mu\nu}(x-y)
\equiv \langle{\rm \tilde T}{\bm A}^a_{\mu}(x){\bm A}^b_{\nu}(y)\rangle, \\
G^>_{ab,\mu\nu}(x-y)
\equiv \langle {\bm A}^a_{\mu}(x){\bm A}^b_{\nu}(y)\rangle, && \
G^<_{ab,\mu\nu}(x-y)
\equiv \langle {\bm A}^b_{\nu}(y){\bm A}^a_{\mu}(x)\rangle.
\end{eqnarray}

In leading-order of perturbative expansion, we need to evaluate the two-point functions at leading order as well as to truncate the expansion of $S_{\rm FV}$ at the second order.
When we are interested in the long-time dynamics of heavy quarks at long-distance scale, e.g. $|\vec x-\vec y|\sim 1/gT \gg 1/T$,
the interaction of soft gluons and hard particles in the medium must be included by resumming the diagrams that contain the hard thermal loops (the hard thermal loop resummation) in order to yield the consistent leading order perturbation \cite{LeBellacText}.
In such a case, the 4-Fermi terms of heavy quark currents in $S_{\rm FV}$ describe the screening and scattering by the $t$-channel interaction between a heavy quark and a hard medium particle.

Although the argument so far is applicable to heavy fermions (muons) in QED straightforwardly, there are a few differences to be remarked.
In QED without light fermions (electrons), the expansion terminates at the second order in $j^{\mu}_1$ and $j^{\mu}_2$, while in QED with electrons, the expansion continues.
Also, we need to subtract from $j^{\mu=0}_1$ and $j^{\mu=0}_2$ the infinite background charge of heavy fermions which fill the Dirac sea.
In QCD, the expansion continues even without light quarks, while heavy quarks in the Dirac sea do not contribute to the background color charge.

\subsection{Non-relativistic limit}
\label{sec:IF-C}
The large separation of energy scales between the environment and the heavy quark system $M\gg T$ allows us to treat heavy quarks as non-relativistic particles.
In order to describe their dynamics, let us carry out the Foldy-Wouthuysen transformation on the heavy quark fields in momentum space $\psi(t,\vec p)$ \cite{Foldy:1949wa}:
\begin{eqnarray}
\psi'(t,\vec p)=\frac{E_p+M+\vec\gamma\cdot\vec p}{\left\{2E_p(E_p+M)\right\}^{1/2}} \psi(t,\vec p),
\end{eqnarray}
with $\gamma^{\mu}$ being the Dirac matrices in the Dirac representation.
This transformation results in the following mass and kinetic terms expanded in terms of $1/M$:
\begin{eqnarray}
\mathcal{L}_{\psi}&=& \mathcal{L}_{\psi}^{_{\rm NR}} + \mathcal O(T^2/M),\\
\mathcal{L}_{\psi}^{_{\rm NR}}&\equiv&
Q^{\dagger}\left(i\partial_0-M+\frac{\nabla^2}{2M}\right)Q
+Q_c\left(i\partial_0+M-\frac{\nabla^2}{2M}\right)Q_c^{\dagger}.
\end{eqnarray}
Here $Q$ and $Q_c$ are the two-component Pauli spinors for a heavy quark and heavy antiquark, defined by $\psi'= \ ^T(Q,Q_c^{\dagger})$.
We make the parametric estimate $\nabla Q_{(c)}^{(\dagger)}\sim \sqrt{MT}\cdot Q_{(c)}^{(\dagger)}$ since the typical momentum of heavy quarks is $p\sim \sqrt{MT}$ in thermal equilibrium.
In this approximation, the kinetic term in $\mathcal{L}_{\psi}^{_{\rm NR}}$ behaves as $\sim(M+T)$.
The non-relativistic limit of the currents is in turn obtained as
\begin{eqnarray}
j^{a0}&=&Q^{\dagger}t^a Q+Q_c t^a Q_c^{\dagger}, \ \ \ \vec j^{a} = \vec j^{a}_{v}+\vec j^{a}_{m}\\
\vec j^{a}_{v}&=&Q^{\dagger}\left(\frac{\vec\nabla-\overleftarrow\nabla}{2iM}\right)t^a Q
-Q_c\left(\frac{\vec\nabla-\overleftarrow\nabla}{2iM}\right)t^a Q_c^{\dagger}+\mathcal{O}\left((T/M)^{3/2}\right),\\
\vec j^a_{m}&=&Q_c\vec\sigma t^a Q + Q^{\dagger}\vec\sigma t^a Q_c^{\dagger}+\mathcal{O}\left(T/M\right).
\end{eqnarray}
Here $\vec\sigma$ denotes the Pauli matrices.
The vector meson term $\vec j^a_{m}$ represents heavy quark pair creation and annihilation processes through the vector channel.
We assume $\vec j^{a}\approx \vec j^{a}_{v}$ which amounts to eliminating $\vec j^a_{m}$ from the currents, equivalent to making the quenched approximation.
The strength of the currents is then estimated as $j^{a0}\propto 1$ and $\vec j^{a}_{v}\propto \sqrt{T/M}$.
For later purposes in Sec.~\ref{sec:IF-D}, we also take the additional parametric estimate 
$\partial_0j^{a0}=-\vec\nabla\cdot \vec j_{v}^{a}\propto \sqrt{T/M}$, $\partial_0\vec j_{v}^{a}=0$,
based on the free equation of motion \cite{Nambu:1997vt}.

In the heavy-ion collisions, the heavy quarkonia are created through the initial hard partonic processes such as gluon fusion and are annihilated into dileptons by electromagnetic processes.
In these processes, $\vec j^{a}_{m}$ or $\vec j_{m}\equiv Q_c\vec\sigma Q + Q^{\dagger}\vec\sigma Q_c^{\dagger}+\cdots$ give the essential contribution.
Since we concentrate on the dynamics of heavy quarks in the QGP, such creation and electromagnetic annihilation processes can be considered as independent from the in-medium dynamics of heavy quarks and thus are out of the scope of this paper. 
The $\vec j^a_m$ can contribute to the in-medium dynamics of heavy quarks through the processes involving (i) virtual heavy quark pairs, (ii) thermal creation of heavy quark pairs, and (iii) pair annihilation of heavy quarks.
The process (i) is limited to very short time and distance ($\sim 1/M$) due to the high virtuality, probability of (ii) is strongly suppressed by $\sim e^{-2M/T}$, and decay width by the process (iii) in a heavy quark bound state (otherwise heavy quark number density is too dilute for heavy quarks to meet each other) is suppressed by the OZI rule 
\cite{Okubo:1963fa}.

\subsection{Long-time behavior}
\label{sec:IF-D}
Since the time scale of the heavy quark dynamics is expected to be much longer than that of the dynamics of medium particles, let us concentrate on the low frequency behavior of the gluonic two-point functions (for a moment, the superscripts and subscripts are suppressed as much as possible):
\begin{eqnarray}
G(\omega,\vec x-\vec y)&\approx& \bar G(\vec x-\vec y)+\omega\bar G'(\vec x-\vec y),\\
\bar G(\vec x-\vec y)&\equiv& G(\omega=0,\vec x-\vec y)=\int_{-\infty}^{\infty} d(x^0-y^0) G(x-y),\\
\bar G'(\vec x-\vec y)&\equiv&\frac{\partial}{\partial \omega} G(\omega,\vec x-\vec y)\Bigr|_{\omega=0}=i\int_{-\infty}^{\infty} d(x^0-y^0)(x^0-y^0)G(x-y).
\end{eqnarray}
This is equivalent to the instantaneous approximation:
\begin{eqnarray}
G(x-y)&\approx &\bar G(\vec x-\vec y)\delta(x^0-y^0)+i\bar G'(\vec x-\vec y)\frac{\partial}{\partial (x^0-y^0)}\delta(x^0-y^0),
\end{eqnarray}
with which the interaction term is approximated as
\begin{eqnarray}
&&\int d^4xd^4y j(x)G(x-y)j(y)\\
&&\approx
\int dt\int d^3xd^3y \left[
\bar G(\vec x-\vec y)j(t,\vec x)j(t,\vec y)
-\frac{i}{2}\bar G'(\vec x-\vec y)\left\{\partial_0 j(t,\vec x)j(t,\vec y)-j(t,\vec x)\partial_0 j(t,\vec y)\right\}
\right].
\nonumber
\end{eqnarray}
In covariant and Coulomb gauges, $\bar G_{0i}=0$ results from the tensor structures.
Since we want to derive the influence action $S_{\rm FV}\left[j^{a\mu}_1,j^{a\mu}_2\right]$ at leading order in $\mathcal O(g^2)$ and up to $\sqrt{T/M}$, only $\bar G_{00}$ and $\bar G'_{00}$ matter in these gauges.
Note that we have made instantaneous approximation to describe the long-time behavior of heavy quarks and obtained the interaction term in the $1/M$ expansion.
In other words, the validity of the low frequency expansion is justified by the hierarchy of scales $M\gg T$.

\subsection{Effective action}
\label{sec:IF-E}
All of the gluonic two-point functions in the instantaneous approximation are expressed in terms of two real functions, $\bar G^{\rm R}_{ab,00}(\vec x-\vec y)$ and $\bar G^{>}_{ab,00}(\vec x-\vec y)$:
\begin{eqnarray}
\bar G^{\rm F}_{ab,00}(\vec x-\vec y)
&=&\left(\bar G^{\rm \tilde F}_{ab,00}(\vec x-\vec y)\right)^*
=-i\bar G^{\rm R}_{ab,00}(\vec x-\vec y)+\bar G^>_{ab,00}(\vec x-\vec y),\\
\bar G'^{\rm F}_{ab,00}(\vec x-\vec y)
&=&\bar G'^{\rm \tilde F}_{ab,00}(\vec x-\vec y)=0,\\
\bar G^{<}_{ab,00}(\vec x-\vec y)&=&\bar G^{>}_{ab,00}(\vec x-\vec y),\\
\bar G'^{>}_{ab,00}(\vec x-\vec y)&=&-\bar G'^{<}_{ab,00}(\vec x-\vec y)=\frac{1}{2T}\bar G^{>}_{ab,00}(\vec x-\vec y),
\end{eqnarray}
where we define the retarded propagator $G^{\rm R}_{ab,00}(\vec x-\vec y)\equiv i\theta(x^0-y^0)\langle[{\bm A}^a_{0}(x),{\bm A}^b_{0}(y)]\rangle$.
In the instantaneous approximation, $\bar G^{\rm R}_{ab,00}(\vec x-\vec y)$ and $\bar G^{>}_{ab,00}(\vec x-\vec y)$ are expressed in terms of an Euclidean correlator and a spectral function:
\begin{eqnarray}
\bar G^{\rm R}_{ab,00}(\vec x-\vec y)
&=& \int_0^{\beta}d\tau\langle {\bm A}_{a0}(t=-i\tau,\vec x){\bm A}_{b0}(t=0,\vec y,)\rangle,\\
\bar G^>_{ab,00}(\vec x-\vec y)&=&T\frac{\partial}{\partial \omega}
\sigma_{ab,00}(\omega,\vec x-\vec y)\Bigr|_{\omega = 0},\\
\sigma_{ab,00}(\omega,\vec x-\vec y)&\equiv&
\int^{\infty}_{-\infty}dx^0 e^{i\omega(x^0-y^0)}
\langle[{\bm A}_{a0}(x),{\bm A}_{b0}(y)]\rangle.
\end{eqnarray}
The detailed derivation of these relations is given in the Appendix \ref{app:2PF}.

In summary, we have derived an effective action $S_{1+2}$ defined as
\begin{eqnarray}
S_{1+2}
&\equiv& S_{\rm kin}+S_{\rm FV}\nonumber\\
&=&\int d^4x\left[
\mathcal{L}_{\psi}(\bar\psi_1,\psi_1)-\mathcal{L}_{\psi}(\bar\psi_2,\psi_2)
\right]+S_{\rm FV},
\end{eqnarray}
at leading order of the perturbative expansion and in the non-relativistic limit :
\begin{eqnarray}
\label{eq:L12}
S_{1+2}&=& S^{_{\rm NR}}_{\rm kin}+S^{_{\rm LONR}}_{\rm FV}+\cdots\\
S^{_{\rm NR}}_{\rm kin}&=&\int d^4x\left[
\mathcal{L}_{\psi}^{_{\rm NR}}({\psi'_1}^{\dagger},\psi'_1)-\mathcal{L}_{\psi}^{_{\rm NR}}({\psi'_2}^{\dagger},\psi '_2)
\right],\\
\label{eq:LFV_NR2}
S_{\rm FV}^{_{\rm LONR}}&=&-\frac{1}{2}\int dtd^3xd^3y\left[
V(\vec x-\vec y)j^{a0}_{1}(t,\vec x)j^{a0}_{1}(t,\vec y)
-V^*(\vec x-\vec y)j^{a0}_{2}(t,\vec x)j^{a0}_{2}(t,\vec y)\right] \nonumber \\
&& + \ i\int dtd^3xd^3y
D(\vec x-\vec y)
j^{a0}_{1}(t,\vec x)j^{a0}_{2}(t,\vec y)\nonumber\\
&& -\frac{1}{4T}\int dtd^3xd^3y
\vec\nabla_x D(\vec x-\vec y)\cdot \left[
\vec j^{a}_{1,\rm NR}(t,\vec x)j^{a0}_2(t,\vec y)+j^{a0}_1(t,\vec x)\vec j^{a}_{2,\rm NR}(t,\vec y)
\right],
\end{eqnarray}
where $\vec j^{a}_{\rm NR}\equiv Q^{\dagger}\left(\frac{\vec\nabla-\overleftarrow\nabla}{2iM}\right)t^a Q-Q_c\left(\frac{\vec\nabla-\overleftarrow\nabla}{2iM}\right)t^a Q_c^{\dagger}$
\footnote{
In leading order of $\sqrt{T/M}$, alternative (non-hermitian) definitions such as
$\vec j^a_{\rm NR}=
Q^{\dagger}\left(\frac{\vec\nabla}{iM}\right)t^a Q -Q_c\left(\frac{\vec\nabla}{iM}\right)t^a Q_c^{\dagger}$
and
$\vec j^{a}_{\rm NR}= \left(\frac{-\vec\nabla}{iM}Q^{\dagger}\right)t^a Q -\left(\frac{-\vec\nabla}{iM}Q_c\right)t^a Q_c^{\dagger}$
lead to the same result since $\nabla D(x)\sim T\cdot D(x)$ while $\nabla Q^{(\dagger)}_{(c)} (x)\sim \sqrt{MT}\cdot Q^{(\dagger)}_{(c)} (x)$.
}.
Here we define a complex potential $V(\vec x-\vec y)$ and a dissipative coupling $D(\vec x-\vec y)$ through:
\begin{eqnarray}
&&-g^2\left\{\bar G^{\rm R}_{ab,00}(\vec x-\vec y)+i\bar G^{>}_{ab,00}(\vec x-\vec y)\right\}\equiv V(\vec x-\vec y)\delta_{ab},\\ 
&&-g^2\bar G^{>}_{ab,00}(\vec x-\vec y)\equiv D(\vec x-\vec y)\delta_{ab}.
\end{eqnarray}
Note that the retarded propagator $\bar G^{\rm R}_{ab,00}(\vec x-\vec y)$ contributes both to the real part of the potential while the forward correlator $\bar G^{>}_{ab,00}(\vec x-\vec y)$ contributes to the imaginary part of the potential and the dissipative coupling.
Since the heavy quark fields $\bar \psi_1, \psi_1$ and $\bar \psi_2, \psi_2$ are coupled by $D(\vec x-\vec y)$, we can interpret Eq.~\eqref{eq:LFV_NR2} as a theory of two interacting species of heavy quarks.

\section{Renormalized Effective Hamiltonian}
\label{sec:REH}
In the previous section, we have obtained the effective action $S_{1+2}$ at leading order of perturbative expansion and in the non-relativistic limit, by which the generating functional $Z[\eta_1,\bar\eta_1,\eta_2,\bar\eta_2]$ reads:
\begin{eqnarray}
\label{eq:HQGF}
Z[\eta_1,\bar\eta_1,\eta_2,\bar\eta_2]&=&\int \mathcal {D}[\bar \psi_1, \psi_1,\bar \psi_2, \psi_2]
\langle\bar\psi_1(-\infty)|{\bm\rho}_{\rm S}|\psi_2(-\infty)\rangle\\
&&\times \exp\Bigl[iS_{1+2}
+i\int d^4x\left(
\bar \eta_1 \psi_1+\bar \psi_1 \eta_1-\bar \eta_2 \psi_2-\bar \psi_2 \eta_2
\right)\Bigr].\nonumber
\end{eqnarray}
In this section, we will construct a renormalized effective Hamiltonian from the effective action Eq.~\eqref{eq:L12} derived in the previous section.
Although the details will be discussed in Sec.~\ref{sec:RTD}, we can already guess from Eq.~\eqref{eq:HQGF} that the effective Hamiltonian will determine the time evolution of a {\it state} $\langle\bar\psi_1|{\bm\rho}_{\rm S}(t)|\psi_2\rangle$ in the form of functional Schr\"odinger equation (the precise definition of $\bm\rho_{\rm S}(t)$ will be given in the next section).
We use the notation $\psi\equiv \ ^T(Q,Q_c^{\dagger})$ for the 4-component Dirac spinor after Foldy-Wouthuysen transformation.

\subsection{Effective Hamiltonian}
\label{sec:REH-A}
Here we derive the effective Hamiltonian $\bm H_{1+2}$ corresponding to the effective action $S_{1+2}=S_{\rm kin}^{_{\rm NR}}+S_{\rm FV}^{_{\rm LONR}}+\cdots$ for the interacting two species of heavy quarks.
In this procedure, we must take care of the time argument of each heavy quark field since the operators must be time-ordered back in the canonical formulation \cite{Ohnuki:1978jv}.
Although all the kinetic and interaction terms have the same time arguments, originally they were different.
Time arguments in the bilinears forms of heavy quark fields, namely the kinetic terms and currents should be read as $\psi^{\dagger}_1(t+\epsilon/2,\vec x)$, $\psi_1(t-\epsilon/2,\vec x)$, $\psi^{\dagger}_2(t-\epsilon/2,\vec x)$, $\psi_2(t+\epsilon/2,\vec x)$ with $\epsilon > 0$.
The products of currents should be understood as sums of all possible time orderings with proper normalization, symbolically expressed as $\frac{1}{n!}\sum_{\sigma\in S_n}\{ \Pi_{i=1}^n j(t_{\sigma(i)},\vec x_i)\}$ with $S_n$ being the permutation group of $n$ elements and $\sigma$ being one of these permutations, where all the time arguments are $t\approx t_1>t_2>\cdots>t_n\approx t$.
$\mathcal{L}_{\rm FV}^{_{\rm LONR}}$ applies to $n=2$ in this expression.
The ordering in the fermion bilinear derives from the time-evolution of the density matrix $e^{-it{\bm H}_{\rm QCD}}{\bm\rho}_{\rm tot} e^{it{\bm H}_{\rm QCD}}$ in the Schr\"odinger picture and that in the current products derives from the current couplings which are nonlocal in time before making the instantaneous approximation.
In order to simplify the equations, we define a notation ${\rm T}_{\rm sym}$ by ${\rm T}_{\rm sym}\{\Pi_{i=1}^n \bm j(t,\vec x_i)\}\equiv\frac{1}{n!}\sum_{\sigma\in S_n}{\rm T}\{ \Pi_{i=1}^n \bm j(t_{\sigma(i)},\vec x_i)\}$.

Now let us introduce new variables $\tilde \psi_{2}\equiv \ ^T \psi_2^{\dagger}$ or $(\tilde Q_2,\tilde Q_{2c}^{\dagger})\equiv(Q_2^{\dagger},Q_{2c})$, which possess properties that will make proceeding arguments transparent.
For example, the time arguments in the fermion bilinears are $\tilde \psi^{\dagger}_2(t+\epsilon/2,\vec x)$, $\tilde \psi_2(t-\epsilon/2,\vec x)$, which are the same with those of $\psi_1^{\dagger}$ and $\psi_1$.
Taking into account the time arguments, the following effective Hamiltonian ${\bm H}_{1+2}={\bm H}^{_{\rm NR}}_{\rm kin}+{\bm H}^{_{\rm LONR}}_{\rm FV}+\cdots$ is derived, where the kinetic term ${\bm H}^{_{\rm NR}}_{\rm kin}$ is
\begin{eqnarray}
{\bm H}^{_{\rm NR}}_{\rm kin}&=&\int d^3xZ_{M}\left\{{\bm \psi}_1^{\dagger} 
\left(\begin{array}{cc}
M & 0 \\
0 & -M
\end{array}
\right)
{\bm \psi_1}
+{\tilde{\bm\psi}}_2^{\dagger} 
\left(\begin{array}{cc}
M & 0 \\
0 & -M
\end{array}
\right)
{\tilde{\bm\psi}}_2
\right\}\nonumber\\
&&+\int d^3x Z_{\psi}\left\{{\bm\psi}_1^{\dagger} 
\left(\begin{array}{cc}
-\frac{\nabla^2}{2M} & 0 \\
0 & \frac{\nabla^2}{2M}
\end{array}
\right)
{\bm \psi}_1
+{\tilde {\bm\psi}}_2^{\dagger}
\left(\begin{array}{cc}
-\frac{\nabla^2}{2M} & 0 \\
0 & \frac{\nabla^2}{2M}
\end{array}
\right)
{\tilde {\bm\psi}}_2
\right\},
\end{eqnarray}
and the interaction Hamiltonian $\bm H_{\rm FV}^{_{\rm LONR}}$ is given by
\begin{eqnarray}
\label{eq:int_H}
\bm H_{\rm FV}^{_{\rm LONR}}&=&\bm H_{(11)} + \bm H_{(22)} + \bm H_{(12)}, \\
\bm H_{(11)}&=&
\frac{Z_{g}^2}{2}\int d^3xd^3y
V(\vec x-\vec y)
{\rm T_{sym}}\left\{
\bm j^{a0}_{1}(t,\vec x)\bm j^{a0}_{1}(t,\vec y)
\right\},\\
\bm H_{(22)}&=&
-\frac{Z_{g}^2}{2}\int d^3xd^3y 
V^*(\vec x-\vec y)
{\rm T_{sym}}\left\{
\bm j^{a0}_{2}(t,\vec x)\bm j^{a0}_{2}(t,\vec y)
\right\},\\
\bm H_{(12)}&=&
-iZ_{g}^2\int d^3xd^3y 
D(\vec x-\vec y){\rm T_{\rm sym}}\left\{
\bm j^{a0}_{1}(t,\vec x)\bm j^{a0}_{2}(t,\vec y)
\right\}\\
&&+\frac{Z_{g}^2}{4T}\int d^3xd^3y 
\vec\nabla_x D(\vec x-\vec y)\cdot{\rm T_{sym}}\left\{
\vec {\bm j}^{a}_{1,\rm NR}(t,\vec x)\bm j^{a0}_{2}(t,\vec y)
+\bm j^{a0}_{1}(t,\vec x)\vec {\bm j}^{a}_{2,\rm NR}(t,\vec y)
\right\}. \nonumber
\end{eqnarray}
The time-ordered currents are:
\begin{eqnarray}
{\rm T}\bm j^{a0}_{1}&=&\bm \psi_1^{\dagger}t^a\bm \psi_1, \ \ \ 
{\rm T}\vec {\bm j}^{a}_{1,\rm NR}=\bm\psi^{\dagger}_1
\left[t^a \otimes \left(
\begin{array}{cc}
1 & 0 \\
0 & -1
\end{array}
\right)\right]
\left(\frac{\vec\nabla-\overleftarrow\nabla}{2iM}\right)
\bm\psi_1, \\
{\rm T}\bm j^{a0}_{2}&=&-\tilde{\bm \psi}_2^{\dagger}t^{a*} \tilde{\bm \psi}_2, \ \ \ 
{\rm T}\vec {\bm j}^{a}_{2,\rm NR}=\tilde{\bm\psi}^{\dagger}_2
\left[t^{a*} \otimes \left(
\begin{array}{cc}
1 & 0 \\
0 & -1
\end{array}
\right)\right]
\left(\frac{\vec\nabla-\overleftarrow\nabla}{2iM}\right)
\tilde{\bm\psi}_2.
\end{eqnarray}
Here $\bm\psi^{\dagger}$, $\bm\psi$, $M$, and $g$ are renormalized fields and parameters and $Z_M$, $Z_{\psi}$, and $Z_g$ are determined by the renormalization conditions.
To be precise, the interaction Hamiltonian Eq.~\eqref{eq:int_H} corresponds to the ladder approximation in the Bethe-Salpeter equation \cite{Nambu:1997vt}.
Cross ladder contributions are of higher order in $g^2$ and thus can be ignored at leading order in the perturbative expansion
\footnote{
Such contributions can be included systematically by the correspondence:
$\exp \left[-i\bm H_{1+2}t\right]
={\rm T}_{\rm (sym)}\exp \left[-i\int_0^t dt' \left(\bm H^{_{\rm NR}}_{\rm kin}+\bm H^{_{\rm LONR}}_{\rm FV}+\cdots\right)\right]$,
where each order of the expansion of the right hand side can produce terms proportional to $t$.
Thus even if we take into account terms only up to $\bm H^{_{\rm LONR}}_{\rm FV}$, this expansion gives $\bm H_{1+2}$ as an expansion in terms of $g^2$, whose leading order (higher order) corresponds to the ladder (cross ladder) contribution to $\bm H_{1+2}$.
}.

In the functional Schr\"odinger equation to be discussed in Sec.~\ref{sec:RTD}, the time-ordered effective Hamiltonian $\bm H_{1+2}$ provides a convenient way to evolve a {\it state} $\rho_{\rm S}[t,\psi_1^{\dagger},\tilde\psi_2^{\dagger}]\equiv \langle \psi_1^{\dagger}|\bm \rho_{\rm S}(t)|\tilde\psi_2^{\dagger}\rangle$ from an initial condition $\rho_{\rm S}[-\infty,\psi_1^{\dagger},\tilde\psi_2^{\dagger}]
=\langle \psi_1^{\dagger}|\bm \rho_{\rm S}|\tilde\psi_2^{\dagger}\rangle$.
In order to obtain the time evolution of $\rho_{\rm S}[t,Q_{1(c)}^*,\tilde Q_{2(c)}^*]\equiv\langle Q_{1(c)}^{*}|\bm \rho_S(t)|\tilde Q_{2(c)}^*\rangle$, which will turn out to be more desirable, it is convenient to reorder $\bm H_{1+2}$ into a normal-ordered product for $\bm Q^{(\dagger)}_{1(c)}$ and ${\tilde{\bm Q}}^{(\dagger)}_{2(c)}$.

\subsection{Renormalization}
\label{sec:REH-B}
Since the operators in the effective Hamiltonian are time-ordered with respect to infinitesimal time differences, interaction terms are expanded in normal order using Wick's theorem with the following non-zero contractions:
\begin{eqnarray}
:\bm\psi_{1\alpha}(x)\bm\psi_{1\beta}^{\dagger}(y):
 &=& \left\{:{\tilde{\bm\psi}}_{2\alpha}(x){\tilde{\bm\psi}}_{2\beta}^{\dagger}(y):\right\}\times(-1) \nonumber \\
&=& \left[1\otimes\left(
\begin{array}{cc}
\theta (x_0-y_0) & 0 \\
0 & -\theta (y_0-x_0)
\end{array}
\right)\right]_{\alpha\beta}\delta(\vec x-\vec y).
\end{eqnarray}
The minus sign for $:{\tilde{\bm\psi}}_{2\alpha}(x){\tilde{\bm\psi}}_{2\beta}^{\dagger}(y):$ derives from the canonical commutation relation:
\begin{eqnarray}
\{\bm\psi_{2\alpha}(t,\vec x),\bm\psi_{2\beta}^{\dagger}(t,\vec y)\}
=\{{\tilde{\bm\psi}}_{2\alpha}^{\dagger}(t,\vec x),{\tilde{\bm\psi}}_{2\beta}(t,\vec y)\}
=-\delta_{\alpha\beta}\delta(\vec x-\vec y).
\end{eqnarray}
Due to the matrix structure of the Wick contraction, one obtains a finite contribution only if the terms are either time-ordered or anti time-ordered.
Therefore any contraction with a fermion loop vanishes, which results from taking the quenched approximation.
Any contraction of a current operator also vanishes because of ${\rm tr} \ t^a=0$.
(In QED, this contribution is infinitely large because of the filled Dirac sea but is canceled by the background charge.)

The normal-ordered product of the kinetic term is
\begin{eqnarray}
\bm H^{_{\rm NR}}_{\rm kin}&=&\int d^3x Z_{M}M\left\{
\bm Q_1^{\dagger}\bm Q_1
+\bm Q_{1c}^{\dagger}\bm Q_{1c}
+{\tilde {\bm Q}}_2^{\dagger} {\tilde {\bm Q}}_2
+{\tilde {\bm Q}}_{2c}^{\dagger} {\tilde {\bm Q}}_{2c}
\right\}\nonumber\\
&&+\int d^3x Z_{\psi}\left\{
\begin{array}{c}
\bm Q_1^{\dagger}\left(-\frac{\nabla^2}{2M}\right)\bm Q_1
+\bm Q_{1c}^{\dagger}\left(-\frac{\nabla^2}{2M}\right)\bm Q_{1c} \\
+{\tilde {\bm Q}}_2^{\dagger}\left(-\frac{\nabla^2}{2M}\right) {\tilde {\bm Q}}_2
+{\tilde {\bm Q}}_{2c}^{\dagger}\left(-\frac{\nabla^2}{2M}\right){\tilde {\bm Q}}_{2c}
\end{array}
\right\},
\end{eqnarray}
and those for the interaction terms are
\begin{eqnarray}
\bm H_{(11)}&=&
\frac{Z_{g}^2}{2}\int d^3xd^3y
V(\vec x-\vec y)
{\rm N}\left\{
\bm j^{a0}_{1}(\vec x)\bm j^{a0}_{1}(\vec y)
\right\} \nonumber \\
&&+\frac{Z_{g}^2C_{\rm F}}{2}\int d^3x
V(0)(\bm Q_1^{\dagger}\bm Q_1+\bm Q_{1c}^{\dagger}\bm Q_{1c})
 + {\rm const.},\\
\bm H_{(22)}&=&
-\frac{Z_{g}^2}{2}\int d^3xd^3y
V^*(\vec x-\vec y)
{\rm N}\left\{
\bm j^{a0}_{2}(\vec x)\bm j^{a0}_{2}(\vec y)
\right\} \nonumber \\
&&+\frac{Z_{g}^2C_{\rm F}}{2}\int d^3x
V^*(0)({\tilde {\bm Q}}_2^{\dagger} {\tilde {\bm Q}}_2
+{\tilde {\bm Q}}_{2c}^{\dagger}{\tilde {\bm Q}}_{2c}) 
 + {\rm const.},\\
\bm H_{(12)}&=&
-iZ_{g}^2\int d^3xd^3y
D(\vec x-\vec y){\rm N}\left\{
\bm j^{a0}_{1}(\vec x)\bm j^{a0}_{2}(\vec y)
\right\}\nonumber\\
&&+\frac{Z_{g}^2}{4T}\int d^3xd^3y 
\vec\nabla_x D(\vec x-\vec y)\cdot {\rm N}\left\{
\vec {\bm j}^{a}_{1,\rm NR}(\vec x)\bm j^{a0}_{2}(\vec y)
+\bm j^{a0}_{1}(\vec x)\vec {\bm j}^{a}_{2,\rm NR}(\vec y)
\right\},
\end{eqnarray}
where N denotes the operation of taking a normal-ordered product and we use $C_{\rm F}=(N_{\rm c}^2-1)/2N_{\rm c}=4/3$.
The complex potential at the origin $V(0)$ consists of a vacuum part and a finite-temperature part $V(0)=V_{\rm vac}+V_{\rm med}$.
$V_{\rm vac}$ is the vacuum electric potential energy at short distance, which is real and divergent, while $V_{\rm med}$ is the finite-temperature contribution to the potential, which is in general complex and finite.

Since we are interested in a situation where heavy quarks are close to thermal equilibrium with typical momentum $k\sim\sqrt{MT}$ and interact with the medium particles by exchanging soft momentum $k\sim gT$, we set the renormalization scale at $k\sim \sqrt{MT}$ for the heavy quark kinetic term and at $k\sim gT$ for the interaction term.
First let us renormalize the potential.
Actually there is no coupling renormalization since the heavy quark internal lines stem from terms we have neglected: the many-body interactions with more than 2 heavy quarks involved and the cross ladder contributions consisting of 2-body interaction.
Therefore we need to choose $Z_g(T)$ so that it normalizes the vacuum potential $V(\vec r)$ at $T=0$ properly at the renormalization scale $r\sim 1/gT$.
At leading order in the perturbative expansion, we can take $Z_g(T)=g(T)/g_{\rm bare}$ or equivalently $g_{\rm bare}=g(T)$ with $Z_g(T)=1$.
The renormalization condition for the kinetic term requires that at $k\sim \sqrt{MT}$ the effective Hamiltonian describes the non-relativistic heavy quark with current mass $M$ at that scale, which is possible only when $M\gg T$.
To be explicit,
\begin{eqnarray}
Z_{\psi}(T)=1,\ \ \ 
Z_{M}(T)+\frac{Z_g(T)^2C_{\rm F}V_{\rm vac}}{2M}=1.
\end{eqnarray}
The renormalized effective Hamiltonian is finally obtained as
\begin{eqnarray}
\label{eq:ren_eff_H}
\bm H_{1+2}&=&\int d^3x
\left[
\begin{array}{l}
aM\left(
\bm Q_1^{\dagger}\bm Q_1
+\bm Q_{1c}^{\dagger}\bm Q_{1c}
\right)\\
+\left\{\bm Q_1^{\dagger}\left(-\frac{\nabla^2}{2M}\right)\bm Q_1
+\bm Q_{1c}^{\dagger}\left(-\frac{\nabla^2}{2M}\right)\bm Q_{1c}
\right\}
\end{array}
\right]\\
&&+\int d^3x
\left[
\begin{array}{l}
a^*M\left(
{\tilde {\bm Q}}_2^{\dagger}{\tilde {\bm Q}}_2
+{\tilde {\bm Q}}_{2c}^{\dagger} {\tilde {\bm Q}}_{2c}
\right)\\
+\left\{{\tilde {\bm Q}}_2^{\dagger}\left(-\frac{\nabla^2}{2M}\right){\tilde {\bm Q}}_2
+{\tilde {\bm Q}}_{2c}^{\dagger}\left(-\frac{\nabla^2}{2M}\right){\tilde {\bm Q}}_{2c}
\right\}
\end{array}
\right]\nonumber\\
&&+\frac{1}{2}\int d^3xd^3y
\left[
\begin{array}{l}
v(\vec x-\vec y)
{\rm N}\left\{\bm j^{a0}_{1}(\vec x)\bm j^{a0}_{1}(\vec y)\right\}
-v^*(\vec x-\vec y)
{\rm N}\left\{\bm j^{a0}_{2}(\vec x)\bm j^{a0}_{2}(\vec y)\right\}\\
-2id(\vec x-\vec y)
{\rm N}\left\{\bm j^{a0}_{1}(\vec x)\bm j^{a0}_{2}(\vec y)\right\}\\
+\frac{1}{2T}\vec\nabla_x d(\vec x-\vec y)\cdot 
{\rm N}\left\{
\vec {\bm j}^{a}_{1,\rm NR}(\vec x)\bm j^{a0}_{2}(\vec y)
+\bm j^{a0}_{1}(\vec x)\vec {\bm j}^{a}_{2,\rm NR}(\vec y)
\right\}
\end{array}
\right]\nonumber\\
&&+\cdots, \nonumber
\end{eqnarray}
with $a\equiv 1+Z_g^2C_{\rm F}V_{\rm med}/2M$ ($a\equiv 1+Z_e^2V_{\rm med}/2M$ for QED), $v(\vec r)\equiv Z_g^2V(\vec r)$, and $d(\vec r)\equiv Z_g^2D(\vec r)$.
The calculation of the gluonic two-point functions $G^{\rm R}_{00}(\omega,\vec k)$ and $G^{>}_{00}(\omega,\vec k)$ at leading order in the hard thermal loop resummed perturbation theory \cite{LeBellacText} yields (adopting coupling renormalization $g_{\rm bare}=g(T)$ and $Z_g(T)=1$)
\begin{eqnarray}
\label{eq:pert-v}
v(\vec r)
&=&\frac{g(T)^2}{4\pi}
\left\{\frac{e^{-\omega_{\rm D} r}}{r}
-2iT\omega_{\rm D}^2\int^{\infty}_{0}dk
\frac{\sin kr}{r(k^2+\omega_{\rm D}^2)^2}\right\},\\
\label{eq:part-d}
d(\vec r)
&=&{\rm Im}\left[v(\vec r)\right]
=-\frac{g(T)^2\omega_{\rm D}^2T}{2\pi}
\int^{\infty}_{0}dk\frac{\sin kr}{r(k^2+\omega_{\rm D}^2)^2},\\
\label{eq:part-a}
a&=&
1+\frac{C_{\rm F}}{2M}\frac{g(T)^2}{4\pi}(-\omega_{\rm D}-iT),
\end{eqnarray}
where $\omega_{\rm D}^2\equiv g(T)^2T^2(N_{\rm c}+N_{\rm f}/2)/3$ for QCD and $\omega_{\rm D}^2\equiv e(T)^2T^2/3$ for QED.

\section{Real-Time Dynamics}
\label{sec:RTD}
In this section, we first derive the functional Schr\"odinger equation for the reduced density matrix.
We then derive the master equation for the reduced density matrix of heavy quarks and also the time-evolution equation for the forward correlator from the renormalized effective Hamiltonian $\bm H_{1+2}$ obtained in the last section.
A stochastic description equivalent to the real-time dynamics is also provided.

\subsection{Functional Schr\"odinger equation for the reduced density matrix}
\label{sec:RTD-A}
First let us start with a reduced density matrix $\bm\rho_{\rm S}(t)$ in the Schr\"odinger picture:
\begin{eqnarray}
\bm\rho_{\rm S}(t)&\equiv& 
{\rm Tr_E}\left\{\bm U(t,-\infty)\bm\rho_{\rm tot}\bm 
U^{\dagger}(t,-\infty)\right\},\\
\bm U(t_1,t_2)&\equiv& \exp \left\{-i(t_1-t_2)\bm H_{\rm QCD}\right\},
\end{eqnarray}
where $\bm\rho_{\rm tot}=\bm\rho^{\rm eq}_{\rm E}\otimes\bm\rho_{\rm S}$ is the initial density matrix for the total system and $\bm U(t_1,t_2)$ is a time-evolution operator with the QCD Hamiltonian $\bm H_{\rm QCD}$.
Note that the reduced density matrix operates only in the heavy quark Fock space.

Now let us split the path integral in Eq.~\eqref{eq:HQGF} at time $t'$ assuming that the source vanishes at $t<t'$:
\begin{eqnarray}
\label{eq:HQGFsplt}
Z[\eta_1,\bar\eta_1,\eta_2,\bar\eta_2]
&=&\int_{[t\geq t']} \mathcal {D}[\bar \psi_1, \psi_1,\bar \psi_2, \psi_2]
\exp\Bigl[i\int_{t'}^{\infty} d^4x\left({\mathcal L}_{1+2}+
\bar \eta_1 \psi_1+\bar \psi_1 \eta_1-\bar \eta_2 \psi_2-\bar \psi_2 \eta_2
\right)\Bigr]\nonumber\\
&& \times \int_{[t<t']}^{\bar\psi_1,\psi_2(t')} \mathcal {D}[\bar \psi_1, \psi_1,\bar \psi_2, \psi_2]
\langle\bar\psi_1(-\infty)|{\bm\rho}_{\rm S}|\psi_2(-\infty)\rangle
e^{i\int_{-\infty}^{t'} d^4x{\mathcal L}_{1+2}},
\end{eqnarray}
where $\int^{\bar\psi_1,\psi_2(t')}_{[t<t']}$ in the second line denotes path integral with the boundary condition at $t=t'$.
The second line gives a functional of the boundary values $\bar\psi_1(t')$ and $\psi_2(t')$ and can be interpreted as a matrix element of the reduced density matrix $\langle\bar\psi_1(t')|{\bm\rho}_{\rm S}(t')|\psi_2(t')\rangle$.
Therefore after change of variables from $\bar\psi$ to $\psi^{\dagger}$, we obtain
\begin{eqnarray}
&&\langle\psi_1^{\dagger}|{\bm\rho}_{\rm S}(t')|\psi_2\rangle
=\int_{[t<t']}^{\psi_1^{\dagger},\psi_2} \mathcal {D}[\psi_1^{\dagger}, \psi_1, \psi_2^{\dagger}, \psi_2]
\langle\psi_1^{\dagger}(-\infty)|{\bm\rho}_{\rm S}|\psi_2(-\infty)\rangle
e^{i\int_{-\infty}^{t'} d^4x{\mathcal L}_{1+2}}.
\end{eqnarray}
Using the analogous relation between the Schr\"odinger's wave equation and the Feynman's path integral formula, we can derive the time-evolution equation of $\rho_{\rm S}[t,\psi_1^{\dagger},\tilde\psi_2^{\dagger}]\equiv\langle\psi_1^{\dagger}|{\bm\rho}_{\rm S}(t')|\tilde\psi_2^{\dagger}\rangle$ with $\tilde\psi_2^{\dagger}=\psi_2$ as a functional Schr\"odinger equation:
\begin{eqnarray}
\label{eq:Schr_FRDMpsi}
i\frac{\partial}{\partial t}\rho_{\rm S}\left[t,\psi_1^{\dagger},\tilde\psi_2^{\dagger}\right]
=H_{1+2}\left[
{\bm\psi}_1^{\dagger}\to\psi_1^{\dagger},
{\bm\psi}_1\to\frac{\delta}{\delta\psi_1^{\dagger}},
\tilde{\bm\psi}_2^{\dagger}\to\tilde\psi_2^{\dagger},
\tilde{\bm\psi}_2\to-\frac{\delta}{\delta\tilde\psi_2^{\dagger}},
\right]\rho_{\rm S}\left[t,\psi_1^{\dagger},\tilde\psi_2^{\dagger}\right].
\end{eqnarray}

The matrix element of $\bm\rho_{\rm S}(t)$ between the heavy quark coherent states $\langle Q_1^*,Q_{1c}^*|$ and $|\tilde Q_2^*,\tilde Q_{2c}^*\rangle$ is given as a functional of the heavy quark fields by
\begin{eqnarray}
\rho_{\rm S}\left[t,Q^*_1,Q^*_{1c},\tilde Q^*_2,\tilde Q^*_{2c}\right]
&=&\langle Q^*_1,Q^*_{1c}|\bm\rho_{\rm S}(t)| \tilde Q_2^*,\tilde Q_{2c}^*\rangle,\\
\langle Q_1^*,Q_{1c}^*|&\equiv&
\langle \Omega |\exp\left[-\int d^3x \left\{
\bm Q(\vec x)Q^*_1(\vec x)+\bm Q_c(\vec x)Q_{1c}^*(\vec x)
\right\}\right],\\
|\tilde Q_2^*,\tilde Q_{2c}^*\rangle&\equiv&
\exp\left[-\int d^3x \left\{
\tilde Q_2^*(\vec x)\bm Q^{\dagger}(\vec x)+\tilde Q_{2c}^*(\vec x)\bm Q_c^{\dagger}(\vec x)
\right\}\right]|\Omega\rangle.
\end{eqnarray}
Here $|\Omega\rangle$ is the heavy quark vacuum state.
In contrast, the coherent states $|\psi_{1,2}\rangle$ and $\langle\psi^{\dagger}_{1,2}|$ are build on an empty Dirac sea $|0\rangle$ which vanishes with $\bm\psi|0\rangle=0$.
Therefore the representation $\rho_{\rm S}\left[t,Q^*_1,Q^*_{1c},\tilde Q^*_2,\tilde Q^*_{2c}\right]$ is more desirable to describe systems with a few heavy quarks in the medium.
The time-evolution of $\rho_{\rm S}\left[t,Q^*_1,Q^*_{1c},\tilde Q^*_2,\tilde Q^*_{2c}\right]$ is obtained by the functional Schr\"odinger equation:
\begin{eqnarray}
\label{eq:Schr_FRDM}
&&i\frac{\partial}{\partial t}\rho_{\rm S}\left[t,Q^*_1,Q^*_{1c},\tilde Q^*_2,\tilde Q^*_{2c}\right]\nonumber\\
&& \ \ \ =H_{1+2}\left[
\bm Q_{1(c)}^{\dagger}\to Q_{1(c)}^*,
\bm Q_{1(c)}\to\frac{\delta}{\delta Q_{1(c)}^*},
{\tilde {\bm Q}}_{2(c)}^{\dagger}\to\tilde Q_{2(c)}^*,
{\tilde {\bm Q}}_{2(c)}\to-\frac{\delta}{\delta \tilde Q_{2(c)}^*}
\right]\nonumber\\
&& \ \ \ \ \ \ \times\rho_{\rm S}\left[t,Q^*_1,Q^*_{1c},\tilde Q^*_2,\tilde Q^*_{2c}\right].
\end{eqnarray}
Note that all the functional differential operators are already moved to the right in the renormalized effective Hamiltonian in which the operators are in the normal order of $\bm{Q}_{1(c)}^{(\dagger)}$ and $\tilde{\bm Q}_{2(c)}^{(\dagger)}$.
Conversely, the procedures taken in the previous section, i.e. the reordering of operators in the normal order and the renormalization, can also be performed in the functional representation to yield the same result.

\subsection{Master equation for the reduced density matrix}
\label{sec:RTD-B}
By functionally differentiating $\rho_{\rm S}\left[t,Q^*_1,Q^*_{1c},\tilde Q_2^*,\tilde Q_{2c}^*\right]$, we obtain, for example, the reduced density matrix for a single heavy quark system:
\begin{eqnarray}
\rho^{ij}_{Q}(t,\vec x,\vec y)
&=&\langle \vec x,i|\bm \rho_Q(t)|\vec y,j\rangle
\propto \langle \Omega| \bm Q^i(\vec x) \bm\rho_{\rm S}(t)\bm Q^{j\dagger}(\vec y)|\Omega\rangle \nonumber\\
&=&-\frac{\delta}{\delta Q^{i*}_1(\vec x)}\frac{\delta}{\delta \tilde Q_2^{j*}(\vec y)}
\rho_{\rm S}\left[t,Q^*_1,Q^*_{1c},\tilde Q_2^*,\tilde Q_{2c}^*\right]\Big |_{Q_{1(c)}^*=\tilde Q_{2(c)}^*=0},
\end{eqnarray}
and that for a heavy quark-antiquark system:
\begin{eqnarray}
\rho^{ijkl}_{QQ_c}(t,\vec x_1,\vec x_2,\vec y_1,\vec y_2)
&=&\langle \vec x_1,i;\vec x_2,j|\bm \rho_Q(t)|\vec y_1,k;\vec y_2,l\rangle\\
&\propto& \langle \Omega| \bm Q^i(\vec x_1)\bm Q^j_c(\vec x_2) \bm \rho_{\rm S}(t) \bm Q_c^{l\dagger}(\vec y_2)\bm Q^{k\dagger}(\vec y_1)|\Omega\rangle \nonumber\\
&=&\frac{\delta}{\delta Q^{i*}_1(\vec x_1)}\frac{\delta}{\delta Q^{j*}_{1c}(\vec x_2)}
\frac{\delta}{\delta \tilde Q_{2c}^{l*}(\vec y_2)}\frac{\delta}{\delta \tilde Q_2^{k*}(\vec y_1)}
\rho_{\rm S}\left[t,Q^*_1,Q^*_{1c},\tilde Q_2^*,\tilde Q_{2c}^*\right]\Big |_{Q_{1(c)}^*=\tilde Q_{2(c)}^*=0}.\nonumber
\end{eqnarray}
The master equation for each of these reduced density matrices is therefore obtained from the functional Schr\"odinger equation for $\rho_{\rm S}\left[t,Q^*_1,Q^*_{1c},\tilde Q_2^*, \tilde Q_{2c}^*\right]$ in Eq.~\eqref{eq:Schr_FRDM}.

For example, let us examine the real-time dynamics of a single heavy quark system at high temperature.
The master equation for a single heavy quark system reads
\begin{eqnarray}
\label{eq:master1}
i\frac{\partial}{\partial t}\rho^{ij}_{Q}(t,\vec x,\vec y)
&=&\left\{(a-a^*)M+\left(-\frac{\nabla_x^2-\nabla_y^2}{2M}\right)\right\}\rho^{ij}_{Q}(t,\vec x,\vec y)\\
&&+\frac{1}{2}
\left(\delta_{ij}\delta_{kl}-\frac{\delta_{ik}\delta_{jl}}{N_{\rm c}}\right)
\left\{-id(\vec x-\vec y)+\frac{\vec\nabla_x d(\vec x-\vec y)}{4T}\cdot\frac{\vec \nabla_x-\vec \nabla_y}{iM}\right\}
\rho^{kl}_{Q}(t,\vec x,\vec y),\nonumber
\end{eqnarray}
and that for the color-traced reduced density matrix $\rho_{Q}(t,\vec x,\vec y)\equiv \rho^{ii}_{Q}(t,\vec x,\vec y)$ is given by
\begin{eqnarray}
\label{eq:master1_tr}
i\frac{\partial}{\partial t}\rho_{Q}(t,\vec x,\vec y)
&=&\left\{(a-a^*)M+\left(-\frac{\nabla_x^2-\nabla_y^2}{2M}\right)\right\}\rho_{Q}(t,\vec x,\vec y)\nonumber\\
&&+C_{\rm F}\left\{-id(\vec x-\vec y)+\frac{\vec\nabla_x d(\vec x-\vec y)}{4T}\cdot\frac{\vec \nabla_x-\vec \nabla_y}{iM}\right\}\rho_{Q}(t,\vec x,\vec y).
\end{eqnarray}
It is interesting to note that taking short distance limit $d(\vec x)\approx d(\vec 0)+\gamma\vec x^2/2$ in the master equation Eq.~\eqref{eq:master1_tr} reduces it to the well-known master equation of the Caldeira-Leggett model for the quantum Brownian motion \cite{Caldeira:1982iu}.
The unitarity of the master equation Eq.~\eqref{eq:master1_tr} is easily confirmed using $\vec\nabla d(\vec x)|_{x=0}=0$:
\begin{eqnarray}
{\rm Tr}_x\bm\rho_{Q}(t)&\equiv&\int d^3x\langle\vec x|\bm\rho_{Q}(t)|\vec x\rangle=\int d^3x\rho_{Q}(t,\vec x,\vec x),\\
i\frac{d}{dt}{\rm Tr}_x\bm\rho_{Q}(t)
&=&\int d^3xd^3y\delta(\vec x-\vec y)
\left(i\frac{\partial}{\partial t}\rho_{Q}(t,\vec x,\vec y)\right)\nonumber\\
&=&\left\{(a-a^*)M-iC_{\rm F}d(\vec 0)\right\}{\rm Tr}_x\bm\rho_{Q}(t)=0.
\end{eqnarray}
Using the master equation Eq.~\eqref{eq:master1_tr}, the following Ehrenfest relations for averaged expectation values 
$\langle \bm{\mathcal{O}}\rangle(t)\equiv {\rm Tr}_x\left\{\bm\rho_{Q}(t)\bm{\mathcal{O}}\right\}=\int d^3xd^3y\langle\vec x|\bm\rho_{Q}(t)|\vec y\rangle\langle\vec y|\bm{\mathcal{O}}|\vec x\rangle$
are derived:
\begin{eqnarray}
\frac{d}{dt}\langle \vec{\bm x}\rangle&=&\frac{\langle \vec{\bm p}\rangle}{M},\\
\frac{d}{dt}\langle \vec{\bm p}\rangle&=&-\frac{\gamma}{2MT}\langle \vec{\bm p}\rangle,\\
\frac{d}{dt}\left\langle \frac{\bm p^2}{2M}\right\rangle
&=&-\frac{\gamma}{MT}\left(\left\langle\frac{\bm p^2}{2M}\right\rangle-\frac{3T}{2}\right).
\end{eqnarray}
To be explicit,
\begin{eqnarray}
\gamma &=&\frac{1}{3}\nabla^2 d(\vec x)|_{x=0}\nonumber\\
&=&-\frac{C_{\rm F}g(T)^2}{9}\nabla^2 \bar G^{>}_{aa,00}(\vec x)|_{x=0}\nonumber\\
&=&\frac{C_{\rm F}g(T)^2}{9}\int\frac{d^3k}{(2\pi)^3}k^2 G^{>}_{aa,00}(\omega=0,\vec k),
\end{eqnarray}
which is consistent with the leading-order perturbative calculation of the drag force \cite{Moore:2004tg}.
In this way, we can derive the consequences of classical Langevin dynamics through the quantum Ehrenfest relations.

\subsection{Time evolution of the forward correlator}
\label{sec:RTD-C}
The forward correlator of a single heavy quark in the Heisenberg picture is defined as
\begin{eqnarray}
G^{>}_{Q,i}(t,\vec x)
&=&{\rm Tr} \left\{{\rm e}^{-\beta \bm H_{\rm QCD}} \bm Q^i(t,\vec x)\bm Q^{j\dagger}(0,\vec 0)\right\}\Big/{\rm Tr}\left\{{\rm e}^{-\beta \bm H_{\rm QCD}}\right\}\nonumber\\
&\approx& \langle\Omega|{\rm Tr}_{\rm E}\left\{\bm\rho_{\rm E} \bm Q^i(t,\vec x)\bm Q^{j\dagger}(0,\vec 0)\right\}|\Omega\rangle,
\end{eqnarray}
where the second equation is a good approximation when $e^{-M/T}\ll 1$.
In the Schr\"odinger picture, the forward correlator is
\footnote{
With an additional assumption $\bm U(t,-\infty)(\bm\rho_{\rm E}^{\rm eq}\otimes\bm\rho_{\rm S})\bm U^{\dagger}(t,-\infty)\sim e^{-\beta\bm H_{\rm QCD}}$ at sufficiently later time $t$, we can show in a similar manner that the reduced density matrix $\rho_{Q}^{ij}(t,\vec x, \vec y)$ converges to the backward correlator $G^{<}_{Q,ij}(t,\vec x, \vec y)={\rm Tr} \left\{{\rm e}^{-\beta \bm H_{\rm QCD}} \bm Q^{i\dagger}(t,\vec x)\bm Q^{j}(t,\vec y)\right\}\Big/{\rm Tr}\left\{{\rm e}^{-\beta \bm H_{\rm QCD}}\right\}$ up to a small multiplication constant $\sim e^{-M/T}$.
The latter defines the Wigner function through a transformation $f(t, \vec x,\vec p)=\int d^3r e^{-i\vec p\cdot\vec r}G^{<}_{Q,ii}(t,\vec x-\vec r/2, \vec x+\vec r/2)$.
}
\begin{eqnarray}
G^{>}_{Q,i}(t,\vec x)
&\approx&\sum_n e^{-\beta E_n}\langle\Omega|\langle n|
\bm U^{\dagger}(t,0)\bm Q^i(\vec x)\bm U(t,0)\bm Q^{j\dagger}(\vec 0)
| n\rangle|\Omega\rangle \Big/ \sum_n e^{-\beta E_n}\nonumber\\
&=&\langle\Omega| \bm Q^i(\vec x){\rm Tr_E}
\left\{
\bm U(t,0)
\left(
\bm\rho_{E}\otimes
\bm\rho_{\rm S}
\right)
\bm U^{\dagger}(t,0)
\right\}
|\Omega\rangle,\\
\bm\rho_{\rm S}&\equiv&\bm Q^{j\dagger}(\vec 0)|\Omega\rangle\langle\Omega|.
\end{eqnarray}
Here we formally take $\bm\rho_{\rm S}=\bm Q^{j\dagger}(\vec 0)|\Omega\rangle\langle\Omega|$ despite the fact that $\bm\rho_{\rm S}$ cannot be understood as a density matrix for any heavy quark system.
This formal treatment is applicable since we did not use any property of the reduced density matrix in deriving the Hamiltonian $H_{1+2}$.
By functional differentiation, we obtain
\begin{eqnarray}
G^{>}_{Q,i}(t,\vec x)&\approx& 
\frac{\delta}{\delta Q^{i*}_1(\vec x)}
\rho_{\rm S}\left[t,Q^*_1,Q^*_{1c},\tilde Q_2^*,\tilde Q_{2c}^*\right]\Big |_{Q_{1(c)}^*=\tilde Q_{2(c)}^*=0},
\end{eqnarray}
with an initial condition $G^{>}_{Q,i}(t=0,\vec x)=\delta_{ij}\delta(\vec x)$.
Similarly, we obtain the forward correlator for the quarkonium state 
\begin{eqnarray}
G^{>}_{QQ_c,ij}(t,\vec x,\vec y ; \vec r)
&=&{\rm Tr} \left\{{\rm e}^{-\beta \bm H_{\rm QCD}} 
\bm Q^i(t,\vec x)\bm Q_c^j(t,\vec y)
\bm Q_c^{l\dagger}(0,-\vec r/2)\bm Q^{k\dagger}(0,\vec r/2)\right\}
\Big/{\rm Tr}\left\{{\rm e}^{-\beta \bm H_{\rm QCD}}\right\}\nonumber\\
&\approx& \langle\Omega|{\rm Tr}_{\rm E}\left\{\bm\rho_{\rm E}
\bm Q^i(t,\vec x)\bm Q_c^j(t,\vec y)
\bm Q_c^{l\dagger}(0,-\vec r/2)\bm Q^{k\dagger}(0,\vec r/2)\right\}|\Omega\rangle\nonumber\\
&=& \frac{\delta}{\delta Q^{i*}_1(\vec x)}
\frac{\delta}{\delta Q^{j*}_{1c}(\vec y)}
\rho_{\rm S}\left[t,Q^*_1,Q^*_{1c},\tilde Q_2^*,\tilde Q_{2c}^*\right]\Big |_{Q_{1(c)}^*=\tilde Q_{2(c)}^*=0},
\end{eqnarray}
with an initial condition $G^{>}_{QQ_c,ij}(t=0,\vec x,\vec y ; \vec r)=\delta_{ik}\delta_{jl}\delta(\vec x-\vec r/2)\delta(\vec y+\vec r/2)$.

The time-evolution equation for the forward correlator is obtained from that of
$\rho_{\rm S}\left[t,Q^*_1,Q^*_{1c},\tilde Q_2^*, \tilde Q_{2c}^*\right]$ in Eq.~\eqref{eq:Schr_FRDM}.
For example, the time evolution of forward correlators of single heavy quark and quarkonium states is given by
\begin{eqnarray}
i\frac{\partial}{\partial t}G^{>}_{Q,i}(t,\vec x)
&=&\left(-\frac{\nabla^2}{2M}+aM\right)G^{>}_{Q,i}(t,\vec x),\\
i\frac{\partial}{\partial t}G^{>}_{QQ_c,ij}(t,\vec x,\vec y ; \vec r)
&=&\left(-\frac{\nabla_x^2+\nabla_y^2}{2M}+2aM\right)
G^{>}_{QQ_c,ij}(t,\vec x,\vec y ; \vec r)\nonumber \\
&&-\frac{1}{2}\left(
\delta_{ij}\delta_{kl}-\frac{\delta_{ik}\delta_{jl}}{N_{\rm c}}
\right)
v(\vec x-\vec y)G^{>}_{QQ_c,kl}(t,\vec x,\vec y ; \vec r).
\end{eqnarray}
Projecting the heavy quark state on the color singlet $G^{>}_{QQ_c,\bm 1}(t,\vec x,\vec y ; \vec r)\equiv G^{>}_{QQ_c,ii}(t,\vec x,\vec y ; \vec r)$, we get
\begin{eqnarray}
i\frac{\partial}{\partial t}G^{>}_{QQ_c,\bm 1}(t,\vec x,\vec y; \vec r)
&=&\left\{-\frac{\nabla_x^2+\nabla_y^2}{2M}+2aM-C_{\rm F}v(\vec x-\vec y)\right\}
G^{>}_{QQ_c,\bm 1}(t,\vec x,\vec y ; \vec r),
\end{eqnarray}
from which the complex color-singlet potential can be read off as $v_{\bm 1}(\vec r)=2(a-1)M-C_{\rm F}v(\vec r)$.
Using the perturbative results in Eqs.~\eqref{eq:pert-v}, \eqref{eq:part-d}, and \eqref{eq:part-a}, the complex color-singlet potential at leading order in the perturbative expansion is
\begin{eqnarray}
v_{\bm 1}(\vec r)
&=&-\frac{C_{\rm F}g(T)^2}{4\pi}
\left[\omega_{\rm D}+\frac{e^{-\omega_{\rm D} r}}{r}
+iT\left\{1-2\omega_{\rm D}^2\int^{\infty}_{0}dk
\frac{\sin kr}{r(k^2+\omega_{\rm D}^2)^2}\right\}\right],
\end{eqnarray}
which agrees with the complex potential obtained in leading-order perturbation theory \cite{Laine:2006ns,Brambilla:2008cx}.
Projecting onto the color-octet state $G^{>}_{QQ_c,\bm 8}(t,\vec x,\vec y;\vec r)\equiv G^{>}_{QQ_c,ij}(t,\vec x,\vec y;\vec r)-(\delta_{ij}/N_{\rm c})G^{>}_{QQ_c,kk}(t,\vec x,\vec y;\vec r)$, we get
\begin{eqnarray}
i\frac{\partial}{\partial t}G^{>}_{QQ_c,\bm 8}(t,\vec x,\vec y; \vec r)
&=&\left\{-\frac{\nabla_x^2+\nabla_y^2}{2M}+2aM+\frac{v(\vec x-\vec y)}{2N_{\rm c}}\right\}
G^{>}_{QQ_c,\bm 8}(t,\vec x,\vec y ; \vec r).
\end{eqnarray}
The complex color-octet potential $v_{\bm 8}(\vec r)\equiv 2(a-1)M+v(\vec r)/2N_{\rm c}$ is therefore obtained as
\begin{eqnarray}
v_{\bm 8}(\vec r)
&=&\frac{g(T)^2}{4\pi}
\left[-C_{\rm F}\omega_{\rm D}+\frac{1}{2N_{\rm c}}\frac{e^{-\omega_{\rm D} r}}{r}
-iT\left\{C_{\rm F}+\frac{\omega_{\rm D}^2}{N_{\rm c}}\int^{\infty}_{0}dk
\frac{\sin kr}{r(k^2+\omega_{\rm D}^2)^2}\right\}\right],
\end{eqnarray}
in leading-order perturbation theory.
It is interesting to note that the imaginary parts of complex color-singlet and color-octet potentials are different at the origin (${\rm Im}\left[v_{\bm 1}(\vec 0)\right]=0$ and ${\rm Im}\left[v_{\bm 8}(\vec 0)\right]=-T(N_{\rm c}/2)g(T)^2/4\pi$) while they are the same at infinitely long distance ${\rm Im}\left[v_{\bm 1}(\infty)\right]={\rm Im}\left[v_{\bm 8}(\infty)\right]=-TC_{\rm F}g(T)^2/4\pi$.
Since the imaginary part reflects Landau damping due to the collisions between the heavy quark and the medium gluons and light quarks, it is natural to obtain a vanishing (finite) imaginary part in the complex color-singlet (color-octet) potential at the origin.

\subsection{Stochastic description}
\label{sec:RTD-D}
The effective action $S_{\rm 1+2}$ can be expressed partly as a stochastic average using both real ($\xi_{\rm r}^a(x)$) and complex ($\xi^a_1(x),\xi^a_2(x)$) Gaussian white noise
\footnote{
This stochastic expression is not unique.
There is a freedom in how the complex noises couple to the heavy quark fields.
For instance, we can also take $c\cdot\xi^a_1j_1^{a0}$ and $c^{-1}\cdot\xi^{a*}_1\vec\nabla\cdot\vec j^a_{2,{\rm NR}}$ instead of $\xi^a_1j_1^{a0}$ and $\xi^{a*}_1\vec\nabla\cdot\vec j^a_{2,{\rm NR}}$ with some complex coefficient $c\neq 0$.
}:
\begin{eqnarray}
e^{iS_{\rm 1+2}}
&=&e^{iS^{\rm NR}_{\rm kin}+iS^{\rm LONR}_{\rm FV}+\cdots},\\
e^{iS^{\rm LONR}_{\rm FV}}
&=&
\exp\left[-\frac{i}{2}\int dt\int d^3xd^3y 
{\rm Re}\left[V(\vec x-\vec y)\right]
\left(j_1^{a0}(t,\vec x)j_1^{a0}(t,\vec y)-j_2^{a0}(t,\vec x)j_2^{a0}(t,\vec y)\right)\right]\\
&\times&\left\langle \exp\left[-i\int d^4x
\left(
\begin{array}{l}
\left\{\xi_{\rm r}^a(x)+\xi^a_1(x)\right\}j_1^{a0}(x)
-\frac{i}{8T}\xi^a_2(x)\vec\nabla\cdot\vec j_{1,\rm NR}^a(x)\\
-\left\{\xi_{\rm r}^a(x)+\xi^{a*}_2(x)\right\}j_2^{a0}(x)
-\frac{i}{8T}\xi^{a*}_1(x)\vec\nabla\cdot\vec j_{2,\rm NR}^a(x)
\end{array}
\right)\right]\right\rangle_{\xi=(\xi_{\rm r}^a,\xi_1^a,\xi_2^a)},\nonumber
\end{eqnarray}
where all the finite noise correlations are
\begin{eqnarray}
\label{eq:noise_corr}
\langle\xi_{\rm r}^a(t,\vec x)\xi_{\rm r}^b(s,\vec y)\rangle 
&=& -\delta^{ab}\delta(t-s)D(\vec x-\vec y),\\
\langle \xi_i^{a*}(t,\vec x)\xi_j^b(s,\vec y)\rangle 
&=& -\delta^{ab}\delta_{ij}\delta(t-s)2D(\vec x-\vec y).
\end{eqnarray}
Note that $D(\vec r)$ is negative definite so that the probability density for the noise can be normalized.

Following the same procedure as in Sec.~\ref{sec:REH}, the normal-ordered renormalized effective stochastic Hamiltonian $\bm H_{1+2}^{(\xi)}$ is obtained as a sum of $\bm H_{1}^{(\xi)}$ and $\bm H_{2}^{(\xi)}$ which consist only of $\bm Q^{(\dagger)}_{1(c)}$ and $\tilde{\bm Q}^{(\dagger)}_{2(c)}$ respectively:
\begin{eqnarray}
e^{iS^{\rm NR}_{\rm kin}+iS^{\rm LONR}_{\rm FV}}&=&
\left\langle\exp\left[-i\Delta t\sum \bm H^{(\xi)}_{1+2}\right]\right\rangle_{\xi},\ \ \
\bm H_{1+2}^{(\xi)}=
\bm H_{1}^{(\xi)}+\bm H_{2}^{(\xi)},\\ 
\bm H_{1}^{(\xi)}&=&\int d^3x
\left[
\begin{array}{l}
{\rm Re}\left[a\right]M\left(
\bm Q_1^{\dagger}\bm Q_1
+\bm Q_{1c}^{\dagger}\bm Q_{1c}
\right)
+\left\{\bm Q_1^{\dagger}\left(-\frac{\nabla^2}{2M}\right)\bm Q_1
+\bm Q_{1c}^{\dagger}\left(-\frac{\nabla^2}{2M}\right)\bm Q_{1c}
\right\}\\
+\left\{\xi_{\rm r}^a(t,\vec x)+\xi^a_1(t,\vec x)\right\}{\rm N}\left\{ \bm j_1^{a0}(\vec x)\right\}
-\frac{i}{8T}\xi^a_2(t,\vec x)\vec\nabla\cdot {\rm N}\left\{ \vec{\bm j}_{1,\rm NR}^a(\vec x)\right\}
\end{array}
\right]\nonumber\\
&&+\frac{1}{2}\int d^3xd^3y
\left[
\begin{array}{l}
{\rm Re}\left[v(\vec x-\vec y)\right]
{\rm N}\left\{\bm j^{a0}_{1}(\vec x)\bm j^{a0}_{1}(\vec y)\right\}
\end{array}
\right],\\
\bm H_{2}^{(\xi)}&=&\int d^3x
\left[
\begin{array}{l}
{\rm Re}\left[a\right]M\left(
{\tilde {\bm Q}}_2^{\dagger}{\tilde {\bm Q}}_2
+{\tilde {\bm Q}}_{2c}^{\dagger} {\tilde {\bm Q}}_{2c}
\right)
+\left\{{\tilde {\bm Q}}_2^{\dagger}\left(-\frac{\nabla^2}{2M}\right){\tilde {\bm Q}}_2
+{\tilde {\bm Q}}_{2c}^{\dagger}\left(-\frac{\nabla^2}{2M}\right){\tilde {\bm Q}}_{2c}
\right\}\\
-\left\{\xi_{\rm r}^a(t,\vec x)+\xi^{a*}_2(t,\vec x)\right\}{\rm N}\left\{ \bm j_2^{a0}(\vec x)\right\}
-\frac{i}{8T}\xi^{a*}_1(t,\vec x)\vec\nabla\cdot {\rm N}\left\{ \vec{\bm j}_{2,\rm NR}^a(\vec x)\right\}
\end{array}
\right]\nonumber\\
&&-\frac{1}{2}\int d^3xd^3y
\left[
\begin{array}{l}
{\rm Re}\left[v(\vec x-\vec y)\right]
{\rm N}\left\{\bm j^{a0}_{2}(\vec x)\bm j^{a0}_{2}(\vec y)\right\}
\end{array}
\right],
\end{eqnarray}
with renormalized noise correlations
\begin{eqnarray}
\langle\xi_{\rm r}^a(t,\vec x)\xi_{\rm r}^b(s,\vec y)\rangle 
&=& -\delta^{ab}\delta_{ts}d(\vec x-\vec y)/\Delta t,\\
\langle \xi_i^{a*}(t,\vec x)\xi_j^b(s,\vec y)\rangle 
&=& -\delta^{ab}\delta_{ij}\delta_{ts}2d(\vec x-\vec y)/\Delta t.
\end{eqnarray}
The time evolution of $\rho_{\rm S}\left[t,Q^*_1,Q^*_{1c},\tilde Q^*_2,\tilde Q^*_{2c}\right]$ is calculated by
\begin{eqnarray}
\rho_{\rm S}\left[t+\Delta t,Q^*_1,Q^*_{1c},\tilde Q^*_2,\tilde Q^*_{2c}\right]
&=&\left\langle e^{-i\Delta t \bm H_{1+2}^{(\xi)}}\right\rangle_{\xi}
\cdot \rho_{\rm S}\left[t,Q^*_1,Q^*_{1c},\tilde Q^*_2,\tilde Q^*_{2c}\right]\\
&=&\left\langle 1-i\Delta t \bm H_{1+2}^{(\xi)}
-\frac{\Delta t^2}{2}{\bm H_{1+2}^{(\xi)}}^2\right\rangle_{\xi}
\cdot \rho_{\rm S}\left[t,Q^*_1,Q^*_{1c},\tilde Q^*_2,\tilde Q^*_{2c}\right].\nonumber
\end{eqnarray}
Note that stochastic variables are of order ${\mathcal O}(1/\Delta t^{1/2})$ and therefore terms up to ${\bm H_{1+2}^{(\xi)}}^2$ are necessary even in the limit $\Delta t\to 0$.
Since $\bm H_{1+2}^{(\xi)}=\bm H_{1}^{(\xi)}+\bm H_{2}^{(\xi)}$ and $\left[\bm H_{1}^{(\xi)},\bm H_{2}^{(\xi)}\right]=0$, the time-evolution operator is given by
\begin{eqnarray}
&&\left\langle 1-i\Delta t \bm H_{1+2}^{(\xi)}
-\frac{\Delta t^2}{2}{\bm H_{1+2}^{(\xi)}}^2\right\rangle_{\xi}\nonumber\\
&& =\left\langle 
\left\{1-i\Delta t \bm H_{1}^{(\xi)}
-\frac{\Delta t^2}{2} \left\langle{\bm H_{1}^{(\xi)}}^2\right\rangle_{\xi}\right\}
\left\{1-i\Delta t \bm H_{2}^{(\xi)}
-\frac{\Delta t^2}{2} \left\langle{\bm H_{2}^{(\xi)}}^2\right\rangle_{\xi}\right\}
\right\rangle_{\xi},
\end{eqnarray}
in the limit $\Delta t\to 0$
\footnote{
In the stochastic description, the ladder approximation corresponds to taking $\Delta t\to 0$ in the expansion of $e^{-i\Delta t H^{(\xi)}_{1+2}}$.
The cross ladder contributions, which turn out to be of higher order in $g$, are recovered by taking into account the finite correlation time $\Delta t\neq 0$ in the expansion, which is not necessarily an expansion in terms of small {\it dimensionless} number.
See \cite{Akamatsu:2011se} for the similar argument.
}.
Here only the real noise correlation remains finite in $\left\langle{\bm H_{1}^{(\xi)}}^2\right\rangle_{\xi}$ and $\left\langle{\bm H_{2}^{(\xi)}}^2\right\rangle_{\xi}$.
This decomposition enables us to describe the time evolution of the system in terms of a stochastic process with real and complex noises.

For example, the stochastic evolution of a single heavy quark system is obtained as
\begin{eqnarray}
\label{eq:stoch_master}
\rho^{ij}_{Q}(t+\Delta t,\vec x,\vec y)
&=&\left\langle U_{Q,ik}^{(\xi)}(\vec x)\tilde U_{Q,jl}^{(\xi)}(\vec y)\right\rangle_{\xi}\rho^{kl}_{Q}(t,\vec x,\vec y),\\
G^{>}_{Q,i}(t+\Delta t,\vec x)
&=&\left\langle U_{Q,ik}^{(\xi)}(\vec x)\right\rangle_{\xi}G^{>}_{Q,k}(t,\vec x),
\end{eqnarray}
with $U_{Q,ik}^{(\xi)}(\vec x)$ and $\tilde U_{Q,ik}^{(\xi)}(\vec x)$ defined by
\begin{eqnarray}
\frac{\delta}{\delta Q^{i*}_{1}(\vec x)}\left[1-i\Delta t H_{1}^{(\xi)}
-\frac{\Delta t^2}{2} \left\langle{H_{1}^{(\xi)}}^2\right\rangle_{\xi}\right]\Biggl|_{Q_{1(c)}^*=0}
&=&U_{Q,ik}^{(\xi)}(\vec x)\frac{\delta}{\delta Q^{k*}_1(\vec x)},\\
\frac{\delta}{\delta \tilde Q^{j*}_{2}(\vec x)}\left[1-i\Delta t H_{2}^{(\xi)}
-\frac{\Delta t^2}{2} \left\langle{H_{2}^{(\xi)}}^2\right\rangle_{\xi}\right]\Biggl|_{\tilde Q_{2(c)}^*=0}
&=&\tilde U_{Q,jl}^{(\xi)}(\vec x)\frac{\delta}{\delta \tilde Q^{l*}_2(\vec x)}.
\end{eqnarray}
Here $H_{1}^{(\xi)}$ and $H_{2}^{(\xi)}$ are functional operators obtained from $\bm H_{1}^{(\xi)}$ and $\bm H_{2}^{(\xi)}$ by the same substitution as in Eq.~\eqref{eq:Schr_FRDM}.
Explicit matrix forms are
\begin{eqnarray}
U_{Q}^{(\xi)}(\vec x)&=&1-i\Delta t
\left[
aM-\frac{\nabla^2}{2M}+\left\{\xi^a_{\rm r}(x)+\xi^a_1(x)\right\}t^a
+\left\{\vec\nabla\xi^a_2(x)\right\}t^a\frac{\vec\nabla}{8MT}
\right],\\
\tilde U_{Q}^{(\xi)}(\vec x)&=&1+i\Delta t
\left[
a^*M-\frac{\nabla^2}{2M}+\left\{\xi^a_{\rm r}(x)+\xi^{a*}_2(x)\right\}t^{a*}
+\left\{\vec\nabla\xi^{a*}_1(x)\right\}t^{a*}\frac{\vec\nabla}{8MT}
\right].
\end{eqnarray}
Using the time-evolution operators $U^{(\xi)}_{Q,ik}(\vec x)$ and $\tilde U^{(\xi)}_{Q,jl}(\vec x)$, the following stochastic Schr\"odinger equations are obtained:
\begin{eqnarray}
\Psi^i_Q(t+\Delta t,\vec x)&=&U^{(\xi)}_{Q,ik}(\vec x)\Psi^k_Q(t,\vec x),\\
\tilde\Psi^j_Q(t+\Delta t,\vec x)&=&\tilde U^{(\xi)}_{Q,jl}(\vec x)\tilde\Psi^l_Q(t,\vec x),
\end{eqnarray}
with which the density matrix $\rho_Q^{ij}(t,\vec x,\vec y)$ and the forward correlator $G^>_{Q,i}(t,\vec x)$ are given by
\begin{eqnarray}
\rho_Q^{ij}(t,\vec x,\vec y)
&=&\left\langle\Psi^i_Q(t,\vec x)\tilde\Psi^j_Q(t,\vec y)\right\rangle_{\xi},\\
G^>_{Q,i}(t,\vec x)&\propto&\left\langle\Psi^i_Q(t,\vec x)\right\rangle_{\xi}.
\end{eqnarray}
Using this stochastic description and taking the classical limit, we can derive the Langevin equation, which we summarize in the Appendix \ref{app:Langevin}.
A stochastic description for systems with arbitrary number of heavy quarks can also be obtained in a similar manner:
for a single heavy antiquark system, $U^{(\xi)}_{Q_c}(\vec x)$ ($\tilde U^{(\xi)}_{Q_c}(\vec x)$) can be obtained by substituting $-t^{a*}$ for $t^a$ ($t^a$ for $-t^{a*}$) in $U^{(\xi)}_{Q}(\vec x)$ ($\tilde U^{(\xi)}_{Q}(\vec x)$);
for a heavy quark-antiquark system $U^{(\xi)}_{QQ_c}(\vec x, \vec y)=U^{(\xi)}_{Q}(\vec x)+U^{(\xi)}_{Q_c}(\vec y)-1-i\Delta t v(\vec x-\vec y)(t^a)_Q\otimes (-t^{a*})_{Q_c}$ and $\tilde U^{(\xi)}_{QQ_c}(\vec x, \vec y)=\tilde U^{(\xi)}_{Q}(\vec x)+\tilde U^{(\xi)}_{Q_c}(\vec y)-1+i\Delta t v^*(\vec x-\vec y)(-t^{a*})_Q\otimes (t^{a})_{Q_c}$.

It is interesting to observe here that $\tilde U^{(\xi)}_{Q,ij}(\vec x)\neq U^{(\xi)*}_{Q,ij}(\vec x)$ and $\tilde\Psi^i_Q(t,\vec x)\neq\Psi^{i*}_Q(t,\vec x)$.
When $M=\infty$, the complex noises $\xi^a_1(x)$ and $\xi^a_2(x)$ can be eliminated so that $\tilde U^{(\xi)}_{Q,ij}(\vec x)= U^{(\xi)*}_{Q,ij}(\vec x)$ and $\tilde\Psi^i_Q(t,\vec x)=\Psi^{i*}_Q(t,\vec x)$ hold.
This is a purely stochastic process with $U^{(\xi)}_{Q}(\vec x)=\exp[-i\Delta t\xi_{\rm r}^a(x) t^a]$, which is not an irreversible process.
Since the complex noises and the irreversible drag process originate from the same term, i.e. $\left(\vec j^{a}_{1,\rm NR}j^{a0}_2+j^{a0}_1\vec j^{a}_{2,\rm NR}\right)$ in the effective action $S_{1+2}$, it is natural to obtain independent stochastic time-evolution equations for $\Psi^i_Q(t,\vec x)$ and $\tilde\Psi^i_Q(t,\vec x)$ in the presence of the complex noises.
This gives us a theoretical foundation for the stochastic description of Ref.~\cite{Akamatsu:2011se} and also illustrates how it must be extended to the finite heavy quark mass case.
In summary, stochastic processes with the real noise encode only the fluctuations while those with complex noises incorporate the irreversible processes caused by the drag force in the classical Langevin dynamics.

\section{Conclusion and Outlook}
\label{sec:Conclusion}
In this paper, we have studied the real-time dynamics of heavy quarks as an open quantum system in thermal QCD medium.
We base our approach on the closed-time path formalism of non-equilibrium quantum field theory and have worked in leading-order perturbation theory and in the non-relativistic limit to make calculation tractable.
In this approximation, the real-time dynamics of heavy quarks is influenced by the environment through the screening and scattering by the hard medium particles.
We have shown that all time-evolution equations can be derived from the renormalized effective Hamiltonian $\bm H_{1+2}$ for doubled degrees of freedom of heavy quark fields: the master equation for the reduced density matrix $\rho_Q^{ij}(t,\vec x,\vec y)$, the time-evolution equation of forward correlator $G^>_{QQ_c,ij}(t,\vec x,\vec y;\vec r)$, and the stochastic Schr\"odinger equation for forward-propagating ($\Psi^i_Q(t,\vec x)$) and backward-propagating ($\tilde\Psi^i_Q(t,\vec x)$) wave functions.
The renormalized effective Hamiltonian governs the time-evolution of the functional reduced density matrix $\rho_{\rm S} \left[t,Q_1^*,Q_{1c}^*,\tilde{Q}_2^*,\tilde {Q}_{2c}^*\right]$ in the form of a functional Schr\"odinger equation, from which the above time-evolution equations are derived.
In leading-order perturbation theory, the renormalized effective Hamiltonian depends only on two real functions, i.e. the real and imaginary parts of the complex potential $v(\vec r)$.
Note that the imaginary part of the complex potential also determines the drag force in perturbation theory.

In the stochastic description, the real part of the complex potential ${\rm Re}\left[v(\vec r)\right]$ represents the interaction between heavy quarks while the imaginary part of the complex potential $d(\vec r)={\rm Im}\left[v(\vec r)\right]$ determines the real noise correlation.
It is also worth noting that the stochastic description introduces complex noises, whose coupling to heavy quark fields is not uniquely determined.
This indicates that the complex noises are not actual physical observables but only represent mathematically the friction process in the stochastic decomposition.
From a practical point of view, however, introducing complex noises enables us to numerically simulate heavy quark systems, for example a single heavy quark system, by stochastically evolving two 3-dimensional functions $\Psi^i_Q(t,\vec x)$ and $\tilde\Psi^i_Q(t,\vec x)$.
Compared to numerical simulation of a 6-dimensional density matrix $\rho^{ij}_Q(t,\vec x,\vec y)$, the use of complex noises might thus be advantageous.

Although our derivation relies on perturbation theory, it should be helpful to construct phenomenological and effective description in the non-perturbative region, perhaps by modeling the renormalized effective Hamiltonian $\bm H_{1+2}$, from which the various time-evolution equations are derived.
One strategy will be to expand $\bm H_{1+2}$ in terms of heavy quark bilinear $\bm Q_{1(c)}^{\dagger}\bm Q_{1(c)}$ and $\tilde{\bm Q}_{2(c)}^{\dagger}\tilde{\bm Q}_{2(c)}$ with each coefficient to be determined by matching.
Another strategy may be to use the reduced density matrix in thermal equilibrium, which is a static solution of the master equation.
Some information must be gained by comparing the thermodynamic quantities, e.g. the Polyakov loop and its correlator, calculated non-perturbatively and those obtained from the reduced density matrix in equilibrium.
Our analyses have shown that thermal fluctuations and the friction process are both indispensable ingredients to the real-time dynamics of heavy quarks and that the number conservations of heavy quark and heavy antiquark do strictly hold in spite of the complex potential.
In constructing a model, it should also possess these properties.

In order to fully appreciate the non-perturbative dynamics of the deconfinement transition through heavy quark systems, we expect that the next-to-leading order perturbative analysis would be important because of the following reasons.
It enables us to describe (1) the processes involving real gluons and (2) the scatterings of heavy quarks with more than one (virtual or real) gluon.
One of the processes (1) is the so-called gluo-dissociation \cite{Kharzeev:1994pz,Rapp:2009my}, by which heavy quarkonium bound states can be excited to the unbound continuum states.
The processes (2) include transitions between color-singlet heavy quarkonium states, which are important at low temperature, in particular in the confined phase, because only the color-singlet states are relevant there.
Therefore the renormalized effective Hamiltonian at the next-to-leading order will also give us an important insight on how to model these processes in the phenomenological and effective description in the non-perturbative region.

\section*{Acknowledgments}
First of all, I would like to thank Alexander Rothkopf and Chiho Nonaka for carefully reading the manuscript.
I would like to thank Mikko Laine, Alexander Rothkopf, and Yannis Burnier for their kind hospitality at University of Bern and for fruitful discussions.
I am also grateful to Jean-Paul Blaizot, Masayasu Harada, Tetsuo Hatsuda, Sangyong Jeon, Berndt M\"uller, Chiho Nonaka, and Derek Teaney for discussions, advices, and encouragements.
This work was supported by the Sasakawa Scientific Research Grant from the Japan Science Society.

\appendix
\section{Relations among the two-point functions}
\label{app:2PF}
Here we derive some of the relations among the gluonic two-point functions in the instantaneous approximation.
From the definition, it follows that 
\begin{eqnarray}
\bar G^{\rm F}_{ab,00}(\vec x-\vec y)
&=&-i\bar G^{\rm R}_{ab,00}(\vec x-\vec y)
+\bar G^<_{ab,\mu\nu}(\vec x-\vec y),\\
\bar G^{\rm \tilde F}_{ab,00}(\vec x-\vec y)
&=&i\bar G^{\rm R}_{ab,00}(\vec x-\vec y)
+\bar G^>_{ab,\mu\nu}(\vec x-\vec y),\\
\bar G^<_{ab,00}(\vec x-\vec y)&=&\bar G^{>*}_{ab,\mu\nu}(\vec x-\vec y).
\end{eqnarray}
Using the Kubo-Martin-Schwinger (KMS) condition $G^<_{ab,00}(t,\vec x-\vec y)=G^>_{ab,00}(t-i\beta, \vec x-\vec y)$, analyticity of the forward correlator $G^>_{ab,00}(t,\vec x-\vec y)$ in a stripe $-i\beta < x^0-y^0 < 0$ in the complex time plane and assuming that the damping of the correlation at $|x^0-y^0|\to\infty$, the instantaneous approximation of the retarded propagator yields an Euclidean correlator:
\begin{eqnarray}
\bar G^{\rm R}_{ab,00}(\vec x-\vec y)
&=&i\int_{-\infty}^{\infty} dx^0 
\theta(x^0-y^0)\langle[{\bm A}^{a}_{0}(x),{\bm A}^{b}_{0}(y)]\rangle\nonumber\\
&=& \int_0^{\beta}d\tau\langle {\bm A}^{a}_{0}(t=-i\tau,\vec x){\bm A}^{b}_{0}(t=0,\vec y)\rangle.
\end{eqnarray}
Obviously $\bar G^{\rm R}_{ab,00}(\vec x-\vec y)$ is real.
Using the KMS condition, we can derive $\bar G^>_{ab,00}(\vec x-\vec y)=\bar G^<_{ab,00}(\vec x-\vec y)$ and thus $\bar G^>_{ab,00}(\vec x-\vec y)$ is also real.
The above results give the relations among $\bar G_{ab,00}(\vec x-\vec y)$s.

The spectral function of gluon fields, which is a real and odd function of $\omega$, is expressed in terms of the forward correlator and retarded propagator:
\begin{eqnarray}
\sigma_{ab,00}(\omega,\vec x-\vec y)&\equiv&
\int^{\infty}_{-\infty}dx^0 e^{i\omega(x^0-y^0)}
\langle[{\bm A}^a_{0}(x),{\bm A}^b_{0}(y)]\rangle\nonumber\\
&=&(1-e^{-\beta\omega})G^>_{ab,00}(\omega,\vec x-\vec y)\nonumber\\
&=&(e^{\beta\omega}-1)G^<_{ab,00}(\omega,\vec x-\vec y)\nonumber\\
&=&2{\rm Im}\left[G^{\rm R}_{ab,00}(\omega,\vec x-\vec y)\right].
\end{eqnarray}
Using the fact that $\bar G'^{\rm R}_{ab,00}(\vec x-\vec y)$ is purely imaginary, which derives from the definition of $G^{\rm R}_{ab,00}(\omega,\vec x-\vec y)$, we obtain 
\begin{eqnarray}
\bar G^{>}_{ab,00}(\vec x-\vec y)
&=&T\frac{\partial}{\partial \omega}\sigma_{ab,00}(\omega,\vec x-\vec y)\Bigr|_{\omega=0},\\
\bar G'^{>}_{ab,00}(\vec x-\vec y)
&=&-\bar G'^{<}_{ab,00}(\vec x-\vec y)
=-i\bar G'^{\rm R}_{ab,00}(\vec x-\vec y)\nonumber\\
&=&\frac{1}{2}\frac{\partial}{\partial \omega}\sigma_{ab,00}(\omega,\vec x-\vec y)\Bigr|_{\omega=0},
\end{eqnarray}
which completes the derivation of all the relations among two-point functions.

\section{Derivation of the classical Langevin equation}
\label{app:Langevin}
We show here how the classical Langevin equation is derived from the stochastic description.
For simplicity, we derive the Langevin equation for muons in a QED plasma composed of photons and electrons.
Let us start from the stochastic representation of the master equation Eq.~\eqref{eq:stoch_master} adapted to QED:
\begin{eqnarray}
\rho_{Q}(t+\Delta t,\vec x,\vec y)
&=&\left\langle U_{Q}^{(\xi)}(\vec x)\tilde U_{Q}^{(\xi)}(\vec y)\right\rangle_{\xi}\rho_{Q}(t,\vec x,\vec y).
\end{eqnarray}
Using the stochastic evolution operators:
\begin{eqnarray}
U_{Q}^{(\xi)}(\vec x)&=&1-i\Delta t
\left[
aM-\frac{\nabla^2}{2M}+\left\{\xi(x)+\xi_1(x)\right\}
+\left\{\vec\nabla\xi_2(x)\right\}\frac{\vec\nabla}{8MT}
\right],\\
\tilde U_{Q}^{(\xi)}(\vec x)&=&1+i\Delta t
\left[
a^*M-\frac{\nabla^2}{2M}+\left\{\xi(x)+\xi^*_2(x)\right\}
+\left\{\vec\nabla\xi^*_1(x)\right\}\frac{\vec\nabla}{8MT}
\right],
\end{eqnarray}
and taking the noise average only for the complex noises $\xi_1(x)$ and $\xi_2(x)$, the master equation is derived as
\begin{eqnarray}
&&\rho_{Q}(t+\Delta t,\vec x,\vec y)-\rho_{Q}(t,\vec x,\vec y)\nonumber\\
&&=-i\Delta t\left\langle
\begin{array}{l}
(a-a^*)M-\frac{\nabla_x^2-\nabla_y^2}{2M}
+\xi_{\rm r}(x)-\xi_{\rm r}(y)\\
-id(\vec x-\vec y)+\frac{\vec\nabla_x d(\vec x-\vec y)}{4T}\cdot\frac{\vec \nabla_x-\vec 
\nabla_y}{iM}
\end{array}
\right\rangle_{\xi_{\rm r}}
\rho_{Q}(t,\vec x,\vec y).
\end{eqnarray}
Regarding this as a stochastic master equation:
\begin{eqnarray}
&&\rho_{Q}(t+\Delta t,\vec x,\vec y;\xi_{\rm r})-\rho_{Q}(t,\vec x,\vec y;\xi_{\rm r})\nonumber\\
&&=-i\Delta t\left\{
\begin{array}{l}
(a-a^*)M-\frac{\nabla_x^2-\nabla_y^2}{2M}
+\xi_{\rm r}(x)-\xi_{\rm r}(y)\\
-id(\vec x-\vec y)+\frac{\vec\nabla_x d(\vec x-\vec y)}{4T}\cdot\frac{\vec \nabla_x-\vec 
\nabla_y}{iM}
\end{array}
\right\}
\rho_{Q}(t,\vec x,\vec y;\xi_{\rm r}),
\end{eqnarray}
with $\rho_Q(t,\vec x,\vec y)\equiv\langle \rho_{Q}(t,\vec x,\vec y;\xi_{\rm r})\rangle_{\xi_{\rm r}}$,
the following stochastic Ehrenfest relations are derived for each noise history:
\begin{eqnarray}
\langle \vec{\bm x}\rangle(t+\Delta t;\xi_{\rm r})
&=&\langle \vec{\bm x}\rangle(t;\xi_{\rm r})
+\Delta t\frac{\langle \vec{\bm p}\rangle(t;\xi_{\rm r})}{M},\\
\langle \vec{\bm p}\rangle(t+\Delta t;\xi_{\rm r})
&=&\langle \vec{\bm p}\rangle(t;\xi_{\rm r})
-\Delta t\frac{\gamma}{2MT}\langle \vec{\bm p}\rangle(t;\xi_{\rm r})
-\Delta t\int d^3x\vec\nabla\xi_{\rm r}\cdot \rho_Q(t,\vec x,\vec x;\xi_{\rm r}).
\end{eqnarray}
In the classical limit, or more precisely, when the probability density of the muon wave packet is localized around $\vec x\approx\langle \vec{\bm x}\rangle(t;\xi_{\rm r})$ with finite spatial size over which the noise $\xi_{\rm r}(x)$ is effectively constant, we can make an approximation $\rho_Q(t,\vec x,\vec x;\xi_{\rm r})\approx \delta (\vec x-\langle \vec{\bm x}\rangle(t;\xi_{\rm r}))$.
Finally we obtain the classical Langevin equation for momentum diffusion,
\begin{eqnarray}
\langle \vec{\bm p}\rangle(t+\Delta t;\xi_{\rm r})
&=&\langle \vec{\bm p}\rangle(t;\xi_{\rm r})
-\Delta t\frac{\gamma}{2MT}\langle \vec{\bm p}\rangle(t;\xi_{\rm r})
+\Delta t\vec f(\vec x = \langle \vec{\bm x}\rangle(t;\xi_{\rm r})),\\
\vec f(\vec x)&\equiv&-\vec\nabla\xi_{\rm r}(\vec x),\\ 
\langle f_i(\vec x)f_j(\vec x)\rangle
&=&-\frac{\partial}{\partial x_i}\frac{\partial}{\partial y_i}d(\vec x-\vec y)|_{\vec x=\vec y}/\Delta t
=\gamma\delta_{ij}/\Delta t.
\end{eqnarray}

\bibliographystyle{apsrev}

\end{document}